\documentclass[aps,prb,twocolumn,superscriptaddress,showpacs]{revtex4-1}

\usepackage{amssymb}
\usepackage{amsmath}
\usepackage{appendix}
\usepackage{bm}
\usepackage{graphicx}
\usepackage{txfonts}
\usepackage{color}

\newcommand{\bea}{\begin{eqnarray}}
\newcommand{\eea}{\end{eqnarray}}
\newcommand{\bei}{\begin{itemize}}
\newcommand{\eei}{\end{itemize}}
\newcommand{\be}{\begin{equation}}
\newcommand{\ee}{\end{equation}}
\newcommand{\bese}{\begin{subequations}}
\newcommand{\eese}{\end{subequations}}

\newcommand{\bfg}{\begin{figure}}
\newcommand{\efg}{\end{figure}}
\newcommand{\eins}{\mbox{$1 \hspace{-1.0mm} {\bf l}$}}

\newcommand{\smequiv}{\! \equiv \!}
\newcommand{\smeq}{\! = \!}
\newcommand{\smgt}{\! > \!}
\newcommand{\DelR}{\Delta^{\mathrm{R}}}
\newcommand{\DelL}{\Delta^{\mathrm{L}}}
\newcommand{\smgg}{\! \gg \!}

\newcommand{\smlt}{\! < \!}

\newcommand{\smap}{\! \approx \!}
\newcommand{\smneq}{\! \neq \!}
\newcommand{\smpl}{\! + \!}

\newcommand{\smmi}{\! - \!}

\newcommand{\ve}{\varepsilon}

\newcommand{\kt}{k_{\mathrm{B}}T}

\newcommand{\ci}{\mathrm{i}}
\newcommand{\xver}{{\hat{\bf x}}}

\newcommand{\zver}{{\hat{\bf z}}}

\newcommand{\braket}[1]{\left<#1\right>}
\newcommand{\parente}[1]{\left(#1\right)}
\newcommand{\ketLR}[1]{\left|#1\right>}
\newcommand{\ketNR}[1]{\left.#1\right>}
\newcommand{\braLR}[1]{\left<#1\right|}

\newcommand{\ketLRr}[1]{\left|\widetilde{#1}\right>}

\newcommand{\braLRr}[1]{\left<\widetilde{#1}\right|}
\newcommand{\braLNr}[1]{\left<\widetilde{#1}\right.}

\newcommand{\modulo}[1]{\left|#1\right|}

\newcommand{\carbon}{$^{13}$C~}

\begin{document}
\title{Dephasing and Hyperfine Interaction in Carbon Nanotubes\\ Double Quantum Dots: The Disordered Case}
\author{Andres A. Reynoso}
\affiliation{Centre for Engineered Quantum Systems, School of Physics, The University of Sydney, NSW 2006, Australia}
\affiliation{Niels Bohr International Academy, Niels Bohr
Institute, Blegdamsvej 17, 2100 Copenhagen \O, Denmark}
\affiliation{Niels Bohr Institute \& Nano-Science Center, University of Copenhagen, Universitetsparken 5, 2100 Copenhagen, Denmark}
\author{Karsten Flensberg}
\affiliation{Niels Bohr Institute \& Nano-Science Center, University of Copenhagen, Universitetsparken 5, 2100 Copenhagen, Denmark}
\date{May 18, 2012}
\begin{abstract}
We study theoretically the \emph{return probability experiment}, which is used to measure the dephasing time, $T_2^*$, in a double quantum dot (DQD) in semiconducting carbon nanotubes with spin-orbit coupling and disorder induced valley mixing. Dephasing is due to hyperfine interaction with the spins of the ${}^{13}$C nuclei.
Due to the valley and spin degrees of freedom four bounded states exist
for any given longitudinal mode in the quantum dot. At zero magnetic field the spin-orbit coupling and the valley mixing split those four states into two Kramers doublets. The valley mixing term for a given dot is determined by the intra-dot disorder, this leads to: (i) states in the Kramers doublets belonging to different dots being different, and (ii) nonzero interdot tunneling amplitudes between states belonging to different doublets. We show that these amplitudes give rise to new avoided crossings, as a function of detuning, in the relevant two particle spectrum: mixing and crossings of the two electrons in one dot states, $(0,2)$, with the one electron in each dot configuration, $(1,1)$. In contrast to the clean system, sequences of different Landau-Zener processes affect the separation and joining stages of each single-shot measurement and, even in a spin-orbit dominated situation, they affect the outcome of the measurement in a way that strongly depends on the initial state. We find that a well-defined return probability experiment is realized when, at each single-shot cycle, the (0,2) ground state is prepared. In this case, the probability to return to the (0,2) ground state remains unchanged but the valley mixing increases the saturation value of the measured return probability. Finally, we study the effect of the valley mixing in the high magnetic field limit; for a parallel magnetic field the predictions coincide with these for DQDs in clean nanotubes, whereas the disorder effect is always relevant when the magnetic field is perpendicular to the nanotube axis.
\end{abstract}
\maketitle
\section{Introduction}

Since the Loss and DiVincenzo proposal,\cite{LossD98} electrons confined in quantum dots have become one of the most attractive platforms for realizing qubits in condensed matter systems.\cite{HansonReview} For dots devised in GaAs based two-dimensional electron gases (2DEGs), the spins of the confined electrons are affected by the hyperfine interaction with the nuclear spins---the nuclear spin $I_0\smeq3/2$ is common to the $^{69}$Ga, $^{71}$Ga, and $^{75}$As isotopes. In the pursuit of fault-tolerant quantum computation,\cite{TaylorFaultTolerant2005} this interaction has been recognized as one of the primary sources of dephasing,\cite{KhaetskiiNazarov2000,LossGlazman2002,Johnson2005} and the challenge of avoiding this problem has led to the implementation and design of several techniques (such as dynamic nuclear polarization, Hahn echoes, etc.) with general success.\cite{Foletti2009,DavidPRL2010,Mike2011}

In parallel, there has been an increasing interest in devising quantum dots in systems that can be isotopically purified leaving only spinless isotopes. This is the case of silicon,\cite{CulcerSarmaQD2009,*CulcerSarmaQD2010,Morello2010,*Morello2011} graphene, and carbon nanotubes (CNTs), which is the system studied in this work.
As a carbon based system, CNTs profit from the 99$\%$ natural abundance of the zero nuclear spin isotope, $^{12}$C. In particular, the gap found in semiconducting CNTs allows for confining electrons using gate-defined quantum dots; this has motivated a great deal of experimental,\cite{MikeDavid2006,Kuemmeth2008,ChurchillFlensberg2009,Watson2009,Steele2009,Jespersen2010,Chorley2011,Jespersen2011} and theoretical studies.\cite{McCann2004,Charlier2007,Bulaev2008,Fischer2009,TrauzettelLoss2009,PalyiB09,WunschDQDCNT2009,WunschDQDCNT2010,FlensbergMarcus2010,PalyiB10,WeissFlensberg2010,Palyi2011} In particular, the generality of the presence of the spin-orbit coupling effects in quantum dots, due to the nanotube curvature,\cite{Ando2000,HuertasGB2006,JeongJL2009,Izumida2009,Klinovaja2011} has only recently been experimentally recognized;\cite{Kuemmeth2008,ChurchillFlensberg2009} the effect persists even in disordered quantum dots occupied by hundreds of electrons.\cite{Jespersen2010,Jespersen2011}

\begin{figure}[t]
   \centering
\vspace{-0.5cm}
   \includegraphics[width=.43\textwidth]{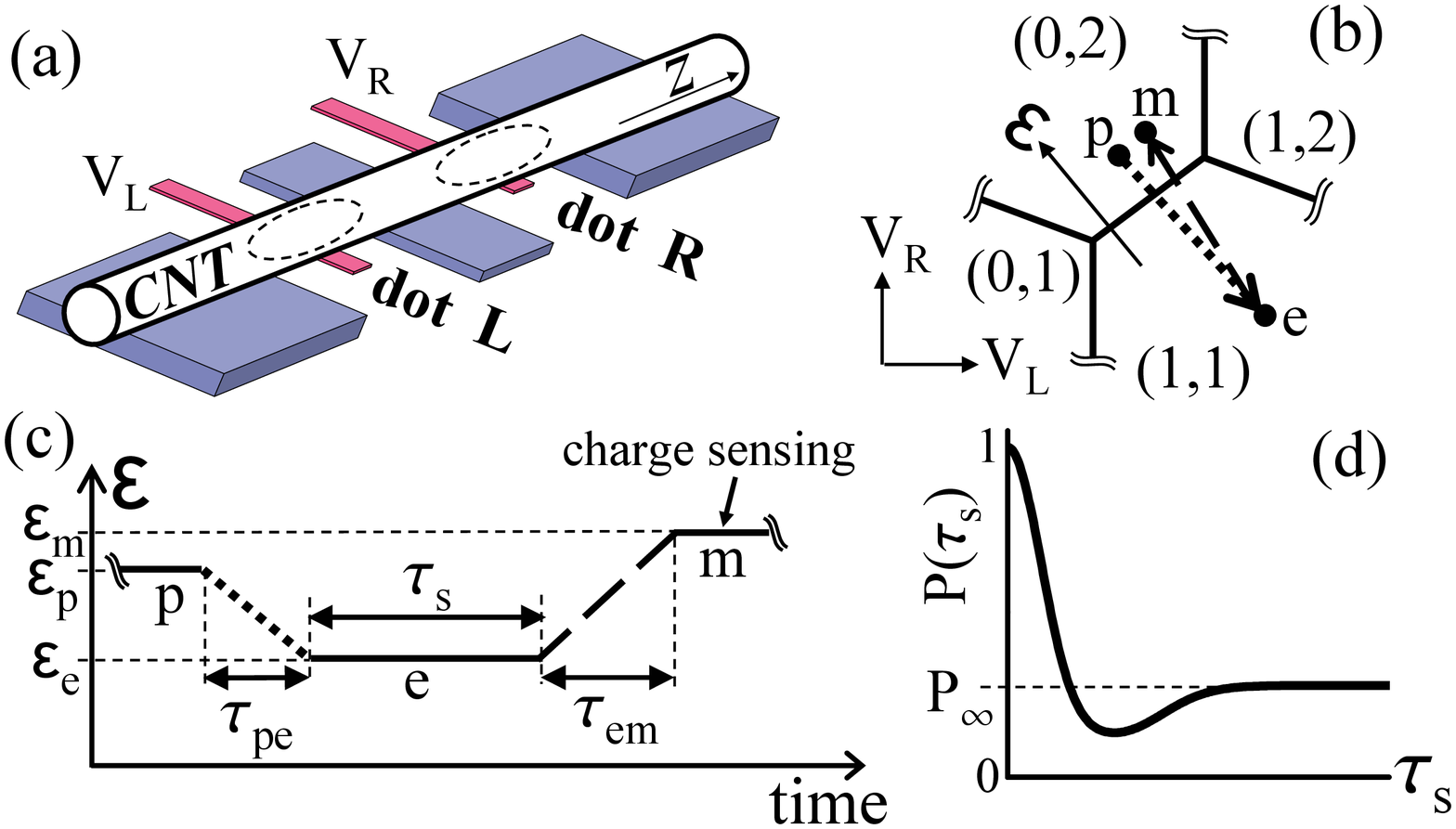}
\vspace{-0.3cm}
   \caption{Double quantum dot in a CNT and the return probability experiment. (a) Scheme of the DQD, gate voltages $V_\mathrm{L}$ and $V_\mathrm{R}$ allow for the control of the number of electrons in each dot. (b) Single-shot experimental cycle in gate voltage space consisting in preparing at point ``p", evolving at point ``e" and measuring at point ``m". The detuning, $\varepsilon$, is zero at the frontier between the $(1,1)$ and $(0,2)$ regions. (c) Simplification of the applied pulse of detuning as a function of time during a single-shot experiment. (d) Sketch of the return probability, $P(\tau_s)$, obtained from averaging many single-shot measurements, the saturation value, $P_\infty$, is qualitatively presented.}
   \label{FG::f1}
\vspace{-0.6cm}
\end{figure}

One of the experimental tools for measuring dephasing is the return probability experiment (RPE), see Fig.\ref{FG::f1}, which involves both control and read-out of the charge state of a double quantum dot, $(N_\mathrm{L},N_\mathrm{R})$, where $N_\mathrm{L}$ and $N_\mathrm{R}$ are the number of electrons in the left and the right dot, respectively. As sketched inside the charge stability map in gate-voltage space in Fig.\ref{FG::f1}(b), each single-shot measurement consists of five stages: (i) \emph{preparation}, the electron is prepared at the $(0,2)$ region; (ii) \emph{separation}, the system is taken adiabatically to the $(1,1)$ region; (iii) \emph{evolution}, electrons are left to evolve under different environments during a time $\tau_s$; (iv) \emph{joining}, the system is taken adiabatically back to the (0,2); and (v) \emph{measuring},  the outcome is nonzero and set to $1$ only if the measured charge state has \emph{returned} to be $(0,2)$. For each $\tau_s$ many single-shot measurements are repeated and their outcomes are averaged leading to the return probability shape, $P(\tau_s)$, that characterizes the dephasing in the system, with a typical $P(\tau_s)$ shape shown in Fig.\ref{FG::f1}(d). Due to this the RPE is also known as \emph{the measurement of $T_2^*$}, which in these systems is the characteristic dephasing time for two-electron states in the double quantum dot. The time scale for performing the fives stages of the RPE should be much shorter than the inelastic relaxation time, $T_1$.

In order to understand the impact of the hyperfine interaction, quantum dots have been devised in nanotubes with natural abundance as well as in samples grown from 99\% $^{13}$C (i.e., nuclear spin $1/2$) enriched methane.\cite{Watson2009,ChurchillFlensberg2009} In Ref. \onlinecite{ChurchillFlensberg2009} the RPE was used to measure $T_2^*$ in a sample rich in $^{13}$C at zero magnetic field, but nor finite magnetic field measurements neither results for samples rich in $^{12}$C are \emph{currently} available.

In absence of defects and impurities, except from the hyperfine interaction effects, there is no scattering between states from inequivalent valleys of the graphene-based band-structure. Already in such a \emph{clean limit} the RPE in a nanotube double dot leads to a rich variety of scenarios. Due to the spin and valley degrees of freedom a CNT DQD can be prepared in six different $(0,2)$ states and, once separated, in the evolution stage, the system have sixteen $(1,1)$ states available for dephasing. The situation is different from GaAs based DQDs in which case only the spin singlet state can be prepared and four $(1,1)$ states are available at the evolution stage.\cite{SchultenWolynes1978,MerkulovHyperfine,CoishDQD,Petta2005,TaylorDephasingTheory2007,Cakr2008,Harju2009} In Ref.~\onlinecite{ReynosoFlensberg2011}, for the RPE in clean nanotubes, we found nine dynamically inequivalent situations and five different lowest bounds for the saturation value of the return probability. The RPE outcome is highly dependent on the prepared state, the curvature induced spin-orbit coupling, and the diamagnetic and the Zeeman effects of an external magnetic field. However, the lowest possible saturation return probability, $P_\infty$ (see Fig.\ref{FG::f1}(d)), is 1/3, as in spin-only DQDs systems, which is above the 0.17 reported in the experiment of Ref.~\onlinecite{ChurchillFlensberg2009}. Therefore, in order to expand our understanding of the RPE to non-ideal samples, we present here a new study that includes a spin-conserving valley mixing interaction which is induced by non-magnetic impurities and defects in the nanotube.\cite{McCannFalko2005}

At zero magnetic field, the valley mixing, which depends on the disorder profile and therefore is different for the two quantum dots, demands us to adopt a more general treatment than the clean case.\cite{ReynosoFlensberg2011} This arises because the products of triplet and singlet functions in valley and spin spaces are no longer the eigenstates of the $(0,2)$ and the $(1,1)$ charge states. Each single dot eigenfunction is a particular combination of the two valleys and the single-particle tunneling amplitudes between solutions of different dots become non-trivial. Therefore, no simple selection rules apply to the mixing between the $(0,2)$ eigenstates and the $(1,1)$ eigenstates, and multiple avoided crossings appear in the two-particle energy versus detuning.

We investigate the physics of the experiment by working in a small tunneling picture that allows us to treat the stages in which the detuning is changed as sequences of different Landau-Zener (LZ) processes. As these concatenated LZ processes affect the separation and the joining stages of the RPE, the outcome of the experiment becomes highly dependent on the shapes of the detuning pulses at those stages---see the arrow starting at point ``p" (``e") and ending at point ``e" (``m") in Fig.\ref{FG::f1}(b) for the separation (joining) stage and the associated shape of the detuning pulse in Fig.\ref{FG::f1}(c). We show that the passage through multiple LZ processes is avoided only if the $(0,2)$ ground state is prepared. The return probability experiment becomes well defined because the preparation guarantees that after the separation stage (at the beginning of the evolution stage) the electrons are in a $(1,1)$ state, i.e., they are really \emph{separated}. In this situation, we find that the saturation return probability, $P_\infty$, is increased above $1/3$; the hyperfine interaction behaves qualitatively as in the clean case whereas the $(1,1)$ \emph{triplet-like} states have an increased probability to return to $(0,2)$ due to a valley-mixing induced avoided crossing that affects the joining stage.

We note that if the highest excited $(0,2)$ state is prepared, $P_\infty\smap 1/6$ is found for a broad region in parameters space. This number is in good agreement with the saturation value reported in the experiment of Ref. \onlinecite{ChurchillFlensberg2009}. However, in our calculations, the multiple LZ processes also affect the separation and joining stages, generating a return probability less than one (even if no dephasing is allowed) by returning to the measurement point without waiting at the $(1,1)$ region, and we find $P_0\smequiv P(\tau_s\smeq 0)\smap 1/2$. Despite the fact that the measurements at short times, $P(\tau_s \rightarrow 0)\smap 1$, are the ones with greatest error bar we conclude that valley mixing does not explain the reported experimental result. In order to draw stronger conclusions more measurements controlling both the prepared state---unknown in Ref.~\onlinecite{ChurchillFlensberg2009}---and the pulse shapes would be desirable.

The behavior for nonzero magnetic fields is also unexplored experimentally. Here we show that, in the limit of high parallel magnetic field, the presence of valley mixing leads to the same predictions obtained for the clean case. This occurs because the diamagnetic effect dominates, so that the single-dot single-particle solutions are valley polarized and the mixing between valleys induced by disorder becomes negligible. On the other hand, a perpendicular external magnetic field, irrespective of its strength, is unable to avoid the effect of the valley mixing because it only introduces Zeeman interaction. Here again, the return probability experiment is not well defined due to the presence of multiple avoided crossings affecting the separation and joining stages. We point out that this highly disorder-dependent mixing between the $(1,1)$ and the $(0,2)$ states in a perpendicular magnetic field can be relevant for understanding Pauli Blockade measurements.

The paper is organized as follows, in the next section we study the model for the single dot including the effective hyperfine field interaction for a Kramers doublet and we introduce the two particle $(0,2)$ and $(1,1)$ eigenstates. We show the effect of interdot tunneling and classify the resulting two-particle spectrum in Section III. Our return probability calculation method and the results are presented in Section IV with conclusions in Section V.

\section{Isolated quantum dots}

In the following, we include Coulomb effects in the constant interaction model, which is valid when the size quantization energy in the quantum dot is large (short dots) or when a strong dielectric substrate screens the interaction,\cite{WunschDQDCNT2009,WunschDQDCNT2010} and thus Wigner molecular states and other interaction effects are not considered. Experimentally, the absence of Coulomb exchange effects has been corroborated in many studies.\cite{Kuemmeth2008,ChurchillFlensberg2009,Jespersen2010,Jespersen2011} Our model is then constructed from two particle Slater determinants based, for convenience, on the single dot eigenfunctions of the left and the right dot. In this section we focus on those single dot wavefunctions, on the effective hyperfine interaction seen by the solutions in a single dot, and we present the two-particle solutions assuming isolated dots.

\subsection{Dot Hamiltonian in valley and spin spaces}

The gate defined quantum dots we study are devised in semiconducting tubes with a bandgap that is due to
either chirality or, for nominally metallic tubes, to curvature.\cite{DresselhausBook,AndoReviewCNT} The low energy $\pi$-band electrons in the CNT are described by two gapped Dirac equations in one spacial dimensions (two Hamiltonian operators acting on four-dimensional spinors due to the spin and pseudo-spin [weight in the two inequivalent sublattices] degrees of freedom), one for each of the two inequivalent valleys in reciprocal space, $K$ and $K'$. Each of the two Dirac equations include the Zeeman interaction, the diamagnetic effect, and \emph{also} spin-orbit coupling terms. For the quantum dots in nanotubes, since the electrostatic confinement potential that defines each dot is smooth on the length scale of the graphene honeycomb lattice spacing, the measured valley mixing effects are not likely to be generated by roughness at interfaces as opposite to other higher dimensional and/or etched-defined systems.\cite{CulcerRoughness2010} Instead, as results in Ref. \onlinecite{McCannFalko2005} suggests, valley mixing in nanotubes are expected as a result of the existence of nonmagnetic impurities and defects within each dot region.

We work with an effective Hamiltonian for the lowest energy bounded electron state. Such a description follows after solving the Dirac equation,\cite{Bulaev2008,WeissFlensberg2010} and has been shown to fit well experimental results.\cite{Kuemmeth2008,Jespersen2010} In contrast to quantum dots in 2DEGs here, due to the coexistence of spin and valley degrees of freedom, the description of a bounded state in the dot is four-dimensional. We introduce the identity and Pauli matrices in spin space $\sigma_i$, with $i\smeq \{0,x,y,z\}$ and the three-dimensional spin vector $\boldsymbol{\sigma}\smequiv(\sigma_x,\sigma_y,\sigma_z)$. The spin projection along the tube axis ($z$-direction) is denoted by $\sigma\!=\{ \uparrow, \downarrow\}$ (or alternatively its numerical version $\sigma \smeq \{+,-\}$). Similarly, the identity and Pauli matrices in valley space are $\tau_j$, with $j\smeq \{0,1,2,3\}$, where we choose $\tau\smeq\{K,K'\}$ as the positive and negative projections of $\tau_3$, respectively; then the three-dimensional valley vector operator is just $\boldsymbol{\tau}\smequiv(\tau_1,\tau_2,\tau_3)$. Neglecting the hyperfine interaction, the description for the right (R) and the left (L) quantum dots is given by the Hamiltonians:
\bea
H^{\xi}_0 &=& \epsilon_{\xi} \sigma_0 \tau_0   -\frac{1}{2} \Delta_{so}^{\xi} \tau_3 \sigma_z
 + \frac{1}{2}\sigma_0 \left\{
\Delta_{KK',1}^{\xi}  \tau_1  + \Delta_{KK',2}^{\xi}  \tau_2\right\} \nonumber\\
&&+\mu_B \left\{  \frac{1}{2} g_{\rm s}  \left({\bf B}\cdot \boldsymbol{\sigma}\right) \tau_0  +  g_{\rm orb}  \left({\bf B}\cdot \zver\right)\sigma_0 \tau_3  \right\},
\label{EQ:Hsingledot}
\eea
with $\xi\smeq\{ {\rm L,R}\}$. The first term describes the effect of the gate voltage applied to the dot $\xi$, we assume that it only introduces a global energy shift of energy $\epsilon_\xi$. The second term, is the spin-orbit coupling splitting energy $\Delta_{so}^\xi$ between $K\sigma$ and  $K'\overline{\sigma}$ states, where $\overline{\sigma}$ stands for the spin projection opposite to $\sigma$. Note that the curvature does not depend on the position in the nanotube and therefore the spin-orbit splitting seen by the lowest lying state is the same for the two dots, $\Delta_{so}^{\rm L}\smeq\Delta_{so}^{\rm R}$, if the dots have the same length.\cite{Jespersen2010} The third term describes the valley mixing in the dot $\xi$. For convenience we work with a single energy parameter (and its phase) to quantify the valley mixing:
\be
\Delta_{\tiny KK'}^{\xi} \equiv \left|\Delta_{KK',1}^{\xi}-\ci \Delta_{KK',2}^{\xi} \right|~,~~
\varphi_{\tiny KK'}^{\xi} \equiv  {\rm Arg} \left\{\Delta_{KK',1}^{\xi}-\ci \Delta_{KK',2}^{\xi} \right\}.
\ee
Since in the RPE the sample under study has its own static pattern of impurities and defects, it is reasonable to assume that $\Delta_{KK'}^\xi$ and $\varphi_{KK'}^\xi$ are fixed sample-specific quantities. Finally, the fourth term describes the effect of an external magnetic field ${\bf B}$, it includes both the usual Zeeman interaction ($\propto g_{\rm s}$) and the diamagnetic effect ($\propto g_{\rm orb}$) which only appears if the external magnetic field has a non-zero component along the tube's axis. Typically the orbital gyromagnetic factor $g_{\rm orb}$ is greater than the spin gyromagnetic factor $g_{\rm s}$ leading to anisotropic magnetic effects.\cite{Minot2004,Kuemmeth2008,Jespersen2010} It is useful to define the Zeeman and orbital splitting energies as $E_{\rm s} \smequiv  g_{\rm s}  \mu_B |{\bf B}|$, and $E_{\rm orb} \smequiv 2 g_{\rm orb} \mu_B \left({\bf B}\cdot \zver\right)$, respectively.

For zero magnetic field (and for magnetic fields along the tube axis) the spin projection $\sigma$ is a good quantum number, then we decompose the Hamiltonian as, $H^{\xi}_0\smeq   H_{\uparrow}^{\xi} + H_{\downarrow}^{\xi}$, with,
\be
  H_{\sigma}^{\xi}= E_{\sigma}^{\xi} \tau_0 +  ({\Delta}^{\xi}_{\sigma}/2) {\hat{\varsigma}^{\xi}_{\sigma}}\cdot \boldsymbol{\tau},
\label{EQ:Hsingledot2by2}
\ee
where,
\begin{subequations}
\bea
E_{\sigma}^{\xi}\equiv \epsilon_{\xi} + \sigma E_{\rm s}/2~,~~~~~\left|{\hat{\varsigma}^{\xi}_{\sigma}}\right|=1, ~~~~~~~~\label{EQ:deltaSigma}
\\  {\hat{\varsigma}^{\xi}_{\sigma}} \equiv \frac{1}{\Delta^{\xi}_{\sigma}}\left(\Delta_{KK'}^{\xi} \cos{\varphi_{\tiny KK'}^{\xi}}, -\Delta_{KK'}^{\xi}\sin{\varphi_{\tiny KK'}^{\xi}},\delta_{\sigma}^{\xi} \right), \label{EQ:valleyDirection}~~~~~\\
\delta_{\sigma}^{\xi}\equiv  E_{\rm orb}- \sigma \Delta_{so}^{\xi} ~,~~~~   {\Delta}^{\xi}_{\sigma}\equiv \sqrt{\left(\Delta_{KK'}^{\xi}\right)^2 + \left(\delta_{\sigma}^{\xi}\vphantom{\Delta_{KK'}^{\xi}}\right)^2 }.~~~
 \label{EQ:isospinfield}
\eea
\label{EQ:gen4}
\end{subequations}

\begin{figure}[t]
     \includegraphics[width=.42\textwidth]{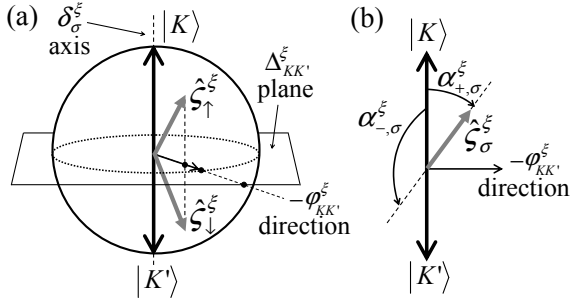}
   \caption{Representation of the Hamiltonians $H_{\sigma}^{\xi}$ and their solutions. (a) Effective Zeeman-like field in the valley space Bloch sphere. The valley mixing is spin independent and it introduces the in-plane component; on the other hand the diamagnetic effect and the spin-orbit coupling enter in the out of plane component. The situation plotted corresponds to zero magnetic field. (b) Angles of the valley-space spinors corresponding to the two eigenstates with spin $\sigma$.}
   \label{FG:Bloch}
\end{figure}

Note that for a given $\sigma$, the 2 by 2 Hamiltonian of Eq. \eqref{EQ:Hsingledot2by2} can be interpreted as a pseudo-Zeeman interaction in valley space. While the spin part of the solution is simply $\ketLR{\sigma}$, the valley component of the eigenstates are the \emph{up} and \emph{down} eigenstates of the valley operator, ${\hat{\varsigma}^{\xi}_{\sigma}}\cdot \boldsymbol{\tau}$. Then the four eigenfunctions of $H^{\xi}_0$, with energies $E_{\xi,\pm,\sigma}\smeq E_{\sigma}^{\xi} \pm \Delta_{\sigma}^{\xi}/2$, are,
\be
\left|\xi,\pm,\sigma\right>=\ketLR{\hat{\varsigma}^{\xi}_{\sigma},\pm} \otimes  \left|\sigma \right>~\mathrm{with}~\ketLR{\hat{\varsigma}^{\xi}_{\sigma},\pm}\equiv\left( \begin{array}{c}
{\rm e}^{\frac{\ci}{2} \varphi_{\tiny KK'}^{\xi}}\cos \frac{\alpha^{\xi}_{\pm,\sigma}}{2} \\
\pm{\rm e}^{-\frac{\ci}{2} \varphi_{\tiny KK'}^{\xi}} \sin\frac{ \alpha^{\xi}_{\pm,\sigma}}{2}
\end{array}
\right),
\label{EQ:eigen}
\ee
where the spinor in valley space for the $\left|\xi,\pm,\sigma\right>$ solution have $\alpha^{\xi}_{\pm,\sigma}$ and $-\varphi_{\tiny KK'}^{\xi}$ as inclination and azimuthal angles, respectively, in the valley Bloch sphere (see Fig.\ref{FG:Bloch}). The inclination angles of the $(+)$ and $(-)$ solutions with spin $\sigma$ fulfill
\begin{subequations}
\bea
\alpha^{\xi}_{+,\sigma}+\alpha^{\xi}_{-,\sigma}&=&\pi, \\
\alpha^{\xi}_{+,\sigma}&\equiv& \arctan\left(\Delta_{\tiny KK'}^{\xi}/\delta_{\sigma}^{\xi}\right)\!\! \mod \pi,
\eea
\label{EQ:angles}
\end{subequations}
which follow from Eq. \eqref{EQ:valleyDirection}.

We see from the expression Eqs. \eqref{EQ:gen4} that when the spin-orbit coupling dominates, the $z$ components of the Zeeman-like valley field, $\delta_{\sigma}^{\xi}$, \emph{have opposite sign} for the two spin projections. This situation is sketched in Fig.\ref{FG:Bloch}(a), the unit vectors $\hat{\varsigma}_{\sigma}^{\xi}$ are drawn on a Bloch sphere in valley space for the ${\bf B}\smeq 0$ case. Focusing on the two lowest (or the two highest) energy eigenfunctions, this implies that solutions for opposite spins have inverted weights in $K$ and $K'$ valleys. On the other hand, if $|E_{\rm orb}| \smgg |\Delta_{so}^{\xi}|$, the sign of $\delta_{\sigma}^{\xi}$ does not depend on $\sigma$. Then, in a high parallel magnetic field limit, $|\delta_{\sigma}^{\xi}|\smgg\Delta_{\tiny KK'}^{\xi}$, and the valley characteristic of the lowest energy solutions for different spin $\sigma$ tends to be the same, i.e., they become valley polarized.

It is worth noting that the Hamiltonians for opposite spin projections are related by the transformation:
\be
 H_{\sigma}^{\xi}(B_z)= \tau_1 \left(H_{\overline{\sigma}}^{\xi}(-B_z)\right)^*  \tau_1,
\label{EQ:sim0}
\ee
which for $B_z\smeq 0$ reduces to the time reversal symmetry relation. The last expression is useful for a direct construction of solutions for opposite spin projections (and opposite field $B_z$) just by interchanging $K$ and $K'$ and conjugating the wavefunctions. Explicitly, this means that for a known solution, $\left|\Psi(\sigma,B_z)\right>$, there exists a related solution with the same energy, $\left|\Psi(\overline{\sigma},-B_z)\right>$, such that,
\bea
\left|\Psi(\sigma,B_z)\right> &=& \psi_{K\sigma} \left|K\sigma\right>
+\psi_{K'\sigma} \left|K'\sigma\right>,\nonumber \\ \left|\Psi(\overline{\sigma},-B_z)\right> &=& \psi_{K'\sigma}^* \left|K\overline{\sigma}\right>
+\psi_{K\sigma}^* \left|K'\overline{\sigma}\right>,
\label{EQ:TRS}
\eea
which in terms of the inclination angles it reduces to recognizing that $\alpha^{\xi}_{\pm,\sigma}(B_z)=\alpha^{\xi}_{\mp,\overline{\sigma}}(-B_z)$. For ${\bf B}\smeq 0$ the two degenerated solutions $\ketLR{\xi,\mathrm{d},\uparrow}$ and $\ketLR{\xi,\mathrm{d},\downarrow}$ fulfill Eq. \eqref{EQ:TRS}. To simplify the notation, we introduce in the following the doublet index, $\mathrm{d}\smeq\{+,-\}$. The two states in the doublet $\mathrm{d}$ make \emph{a Kramers doublet} because they are linked by time reversal symmetry. Each doublet can be regarded as a spin-$\frac{1}{2}$ system where, unlike spin-only quantum dots, the orbital parts of the two states are different.

\subsection{Effective hyperfine interaction inside the doublets}
\label{SC:hyperfine}

For brevity we omit in this section the dot index $\xi \smeq\{ \mathrm{L, R}\}$---for instance the eigenstates $\left|\xi,\mathrm{d},\sigma\right>$ become $\left|\mathrm{d},\sigma\right>$, and the following applies for both dots. The effective Hamiltonian seen by a confined electron in a quantum dot, as a result of the hyperfine interaction with the \carbon nuclear spins, can be cast as,\cite{PalyiB09}
\be
H_{\tiny\rm{HF}}^{\rm eff}= \frac{1}{2}\sum_{i = 0}^2 {\rm \tau}_i {\bf h}^{(i)} \cdot \boldsymbol{\sigma},
\label{EQ:HHFI}
\ee
where we have adopted an isotropic hyperfine field interaction (HFI), although there is some degree of anisotropy.\cite{Fischer2009} The dynamics of the electron spin is much faster than the precession time of the nuclear spins, $T_{nuc}$, and furthermore, each single-shot measurement in the RPE is realized over a time much shorter than $T_{nuc}$. Therefore during each experiment the fields, ${\bf h}^{(i)}$, are constants given by the matrix elements of the hyperfine interaction Hamiltonian, $H_{\tiny\rm{HF}}=A_\mathrm{iso}\sum_{l} \boldsymbol{\sigma} \cdot {\bf I}_{l} \delta({\bf r}-{\bf R}_{l})$, [where $A_\mathrm{iso}$ is the hyperfine coupling constant and the summation is taken over all the lattice sites in the quantum dot region, $l$, that have a \carbon atom with ${\bf I}_{l}$ the spin-$1/2$ vector operator of the nucleus located at ${\bf R}_{l}$] and tracing over the ensemble of nuclear spins.\cite{MerkulovHyperfine}

To simulate the RPE many realizations of the single-shot measurement are averaged over. The numbers $h_j^{(i)}$, with $i\smeq \{0,1,2\}$ and $j\smeq \{x,y,z\}$, follow Gaussian probability distributions with zero mean and the following variances,\cite{PalyiB09}
\be
\sigma_{\rm H}^2=A_{\rm iso}^2 \nu/(4 N_{QD})=\braket{\left({h}^{(0)}_j\right)^2}=
2\braket{\left({h}^{(1)}_j\right)^2}=
2\braket{\left({h}^{(2)}_j\right)^2},~~
\label{EQ:variances}
\ee
where $N_{QD}$ the number of atoms in the quantum dot and $\nu$ is the abundance of \carbon atoms in the dot.

We now focus on dots that have both nonzero $\Delta_{\tiny KK'}$ and nonzero $\Delta_{so}$ splittings.\cite{Kuemmeth2008,ChurchillFlensberg2009,Jespersen2010,Jespersen2011} The estimation for variances in Eq. \eqref{EQ:variances} are typically much
smaller than the energy separation, $\Delta\equiv\sqrt{\left(\Delta_{\tiny KK'}\right)^2+\left(\Delta_{so}\right)^2}$, between the $\mathrm{d}=+$ and the $\mathrm{d}=-$ doublets. Due to this condition, the HFI mixing of electron states belonging to different Kramers doublets can be neglected and therefore we only include matrix elements within each doublet $\mathrm{d}$,
\be
\left<\mathrm{d},\sigma\right| H_{\rm HF}^{\rm eff} \left|\mathrm{d},\sigma'\right>.
\ee
This allows us to write an effective hyperfine interaction for the doublet $\mathrm{d}$ as:
\be
H^{\mathrm{d}}_{\tiny\rm{HF}} = {\bf B^{\mathrm{d}}} \cdot \boldsymbol{\sigma}^{\mathrm{d}}+B^{\mathrm{d}}_0 \sigma^{\mathrm{d}}_0,
\label{EQ:HhfiKds}
\ee
\begin{widetext}
where the operators $\sigma^{\mathrm{d}}_0$, $\sigma^{\mathrm{d}}_x$, $\sigma^{\mathrm{d}}_y$, and $\sigma^{\mathrm{d}}_z$ are the Pauli matrices operating in the $(\mathrm{d})$ Kramers doublet states and
\bea
 B^{\mathrm{d}}_0&=& \frac{1}{2}\mathrm{d} \left(h_z^{(1)}\cos{\varphi_{\tiny KK'}}-h_z^{(2)}\sin{\varphi_{\tiny KK'}}\right)\left(\sin{\alpha_{\mathrm{d},\uparrow}}-\sin{\alpha_{\mathrm{d},\downarrow}}\right), \nonumber \\
 B^{\mathrm{d}}_x&=&  h_x^{(0)}\cos{\left(\frac{\alpha_{\mathrm{d},\uparrow}-\alpha_{\mathrm{d},\downarrow}}{2}\right)} +\mathrm{d}\left(h_x^{(1)}\cos{\varphi_{\tiny KK'}}- h_x^{(2)}\sin{\varphi_{\tiny KK'}}\right)\sin{\left(\frac{\alpha_{\mathrm{d},\uparrow}+\alpha_{\mathrm{d},\downarrow}}{2}\right)} +\mathrm{d}\left(h_y^{(1)}\sin{\varphi_{\tiny KK'}}+ h_y^{(2)}\cos{\varphi_{\tiny KK'}}\right)\sin{\left(\frac{\alpha_{\mathrm{d},\uparrow}-\alpha_{\mathrm{d},\downarrow}}{2}\right)},  \nonumber \\
 B^{\mathrm{d}}_y&=&  h_y^{(0)}\cos{\left(\frac{\alpha_{\mathrm{d},\uparrow}-\alpha_{\mathrm{d},\downarrow}}{2}\right)} +\mathrm{d}\left(h_y^{(1)}\cos{\varphi_{\tiny KK'}}- h_y^{(2)}\sin{\varphi_{\tiny KK'}}\right)\sin{\left(\frac{\alpha_{\mathrm{d},\uparrow}+\alpha_{\mathrm{d},\downarrow}}{2}\right)} -\mathrm{d}\left(h_x^{(1)}\sin{\varphi_{\tiny KK'}}+ h_x^{(2)}\cos{\varphi_{\tiny KK'}}\right)\sin{\left(\frac{\alpha_{\mathrm{d},\uparrow}-\alpha_{\mathrm{d},\downarrow}}{2}\right)}, \nonumber \\
 B^{\mathrm{d}}_z&=& h_z^{(0)} +\frac{1}{2}\mathrm{d} \left(h_z^{(1)}\cos{\varphi_{\tiny KK'}}-h_z^{(2)}\sin{\varphi_{\tiny KK'}}\right)\left(\sin{\alpha_{\mathrm{d},\uparrow}}+\sin{\alpha_{\mathrm{d},\downarrow}}\right). \label{EQ:effHfine}
\eea
\end{widetext}

These effective fields can be readily written as a function of the parameters of the Hamiltonian by using the expressions given in Eqs. \eqref{EQ:angles}. Furthermore, as discussed above, at zero magnetic field the inclination angles of the two solutions within the same doublet are related by time reversal symmetry, $\alpha_{\mathrm{d},\uparrow} \smpl \alpha_{\mathrm{d},\downarrow}\smeq \pi$ (they are opposite spinors in the valley Bloch sphere), and the expressions become simpler.

\begin{figure}[b]
   \centering
   \includegraphics[width=.35\textwidth]{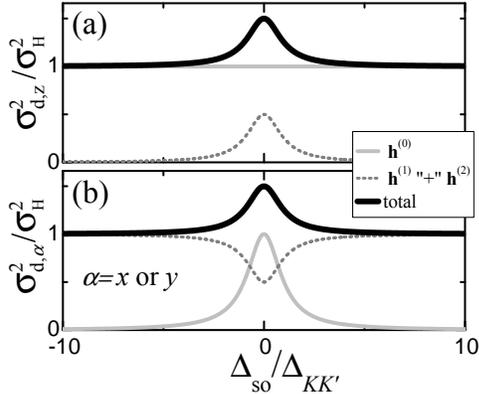}
   \caption{Effect of the valley mixing on the variances of the effective hyperfine fields components ($x$, $y$ and $z$ directions) that act on the subspace of the two states in the Kramers doublet $\mathrm{d}\smeq\{+,-\}$, namely $\ketLR{\mathrm{d},\uparrow}$ and $\ketLR{\mathrm{d},\downarrow}$ (see Eq.\eqref{EQ:eigen}). We also show the partial contributions to the variances that arise as a result of (gray solid line) the valley conserving components of the HFI, $\mathbf{h}^{(0)}$, and (black dotted line) the non-valley conserving HFI components: the addition of the contributions generated by the $\mathbf{h}^{(1)}$ and the $\mathbf{h}^{(2)}$ components.}
   \label{FG:var}
\end{figure}

The new fields also follow Gaussian probability distributions with zero mean. The variances of the effective field for the Kramers doublet $\mathrm{d}$ (defined as $\sigma_{\mathrm{d},\alpha}^2\smequiv\braket{\parente{B^{\mathrm{d}}_{\alpha\vphantom{i}}}^2}$, with $\alpha=0,x,y,z$) are,
\begin{subequations}
\bea
\frac{\sigma_{\mathrm{d},0}^2}{{\sigma_{\rm H}^2}} &=&  \frac{1}{8}\left(\sin^2{\alpha_{\mathrm{d},\uparrow}}+ \sin^2{\alpha_{\mathrm{d},\downarrow}}-2\sin{\alpha_{\mathrm{d},\uparrow}}\sin{\alpha_{\mathrm{d},\downarrow}}\right), \\
\frac{ \sigma_{\mathrm{d},x}^2}{\sigma_{\rm H}^2} &=& \frac{ \sigma_{\mathrm{d},y}^2}{\sigma_{\rm H}^2} = 1 + \frac{1}{2} \sin{\alpha_{\mathrm{d},\uparrow}}\sin{\alpha_{\mathrm{d},\downarrow}},
\eea
\be
\frac{\sigma_{\mathrm{d},z}^2 }{\sigma_{\rm H}^2} = 1 + \frac{1}{8}\left(\sin^2{\alpha_{\mathrm{d},\uparrow}}+ \sin^2{\alpha_{\mathrm{d},\downarrow}}+2\sin{\alpha_{\mathrm{d},\uparrow}}\sin{\alpha_{\mathrm{d},\downarrow}}\right).~~~~~~~
\ee
\label{EQ:variances2}
\end{subequations}

In Fig.\ref{FG:var} we plot the variances of the three components of the effective magnetic field as a function of the ratio $\Delta_{so}/\Delta_{\tiny KK'}$ for the case of zero magnetic field. The component proportional to the identity operator in the Kramers doublet only produces an energy shift and therefore it is irrelevant for the dynamics of the problem. The first important result is that the \emph{total variance} of the three relevant components are equal, moreover, they are the same in the two doublets. This means that the hyperfine field within each doublet is statistically isotropic and therefore the dephasing of a given state in the doublet is similar to the dephasing of an electron spin confined in a GaAs quantum-dot.

We also present the contribution of the valley-mixing terms ($\propto {\bf h}^{(1)}$ and $\propto {\bf h}^{(2)}$) and the valley-conserving terms ($\propto {\bf h}^{(0)}$) of the effective HFI effective Hamiltonian of Eq.\eqref{EQ:HHFI}. Both contributions are isotropic when the disorder dominates. This happens because the eigenstates are valley spinors having equal weight in $\ketLR{K}$ and $\ketLR{K'}$ (the spinors lie in the plane of the valley Bloch sphere) and therefore both classes of terms can mix those states. On the other hand, the states for $\Delta_\mathrm{so}\gg\Delta_{KK'}$ are
\begin{subequations}
\bea
\ketLR{-,\uparrow}&=&\ketLR{K\vphantom{K'}}\otimes\ketLR{\uparrow\vphantom{K'}},~~~\ketLR{-,\downarrow}=\ketLR{K'}\otimes\ketLR{\downarrow\vphantom{K'}}, \\
\ketLR{+,\uparrow}&=&\ketLR{K'}\otimes\ketLR{\uparrow\vphantom{K'}},~~~\ketLR{+,\downarrow}=\ketLR{K\vphantom{K'}}\otimes\ketLR{\downarrow\vphantom{K'}}. \eea
\end{subequations}
In this limit the valley conserving hyperfine operator, $\tau_0 {\bf h}^{(0)}\cdot \boldsymbol{\sigma}$, contributes to $B^{\mathrm{d}}_{z}$ since it provides diagonal matrix elements for $\ketLR{\mathrm{d},\uparrow}$ and $\ketLR{\mathrm{d},\downarrow}$ states; at the same time the operator is unable to flip the valley and it does not contribute to the effective fields $B^{\mathrm{d}}_{x}$ and $B^{\mathrm{d}}_{y}$. On the other hand, the operators $\tau_1 {\bf h}^{(1)}\cdot \boldsymbol{\sigma}$ and $\tau_2 {\bf h}^{(2)}\cdot \boldsymbol{\sigma}$ both flip the valley index and therefore they do not contribute to $B^{\mathrm{d}}_{z}$ but they contribute to $B^{\mathrm{d}}_{x}$ and $B^{\mathrm{d}}_{y}$.

\subsection{The $(0,2)$ and $(1,1)$ eigenstates}

Here we give the two particle solutions for two electrons occupying the right dot and for one electron occupying each in absence of tunneling, i.e., in the high detuning limit valid for the preparation, evolution and measurement stages---see Fig.\ref{FG::f1}(a). For simplicity, we number the four eigenstates and eigenenergies in the dot $\xi\smeq\{\mathrm{L,R}\}$ as $\ketLR{\xi n}$ and $E_{\xi n}$ with $n\smeq \{1,2,3,4\}$. We define a two particle Slater determinant built from states $\xi n$ and $\xi' n'$ as
\begin{equation}
\ketLR{{}_{\xi n}^{\xi' n' }}=
\frac{1}{\sqrt{2}}\left(
\ketLR{\vphantom{\xi'}\xi n}_{\bf
1}\ketLR{\xi' n' }_{\bf
2}-\ketLR{\xi' n'}_{\bf
1}\ketLR{\vphantom{\xi'}\xi n}_{\bf 2} \right).
\label{EQ:Sla0}
\end{equation}

For two electrons in the right dot, the $(0,2)$ charge configuration, only six independent states can be constructed using the four single-particle states $\ketLR{\mathrm{R} n}$. These eigenstates and their eigenenergies are,
\begin{equation}
(0,2):\quad \ketLR{{}_{\mathrm{R}n}^{\mathrm{R}n'}},\quad
E^{(0,2)}_{n,n'}=2\epsilon_\mathrm{R}+E_{\mathrm{R} n}+E_{\mathrm{R}n'} + U_\mathrm{RR},
\label{EQ:02sols0En}
\end{equation}
where $n<n'$ and $U_\mathrm{RR}$ is the Coulomb repulsion energy.

The eigenstates in the $(1,1)$ charge configuration are the Slater determinants constructed from states $\ketLR{\mathrm{R} n}$ and $\ketLR{\mathrm{L} n'}$. Therefore we have the sixteen states and energies:
\begin{equation}
(1,1):\quad \ketLR{{}_{\mathrm{R}n}^{\mathrm{L}n'}},
\quad E^{(1,1)}_{n,n'}=\epsilon_\mathrm{L}+\epsilon_\mathrm{R}+E_{\mathrm{R}n}+E_{\mathrm{L}n'}+U_\mathrm{LR},
\label{EQ:11sols0}
\end{equation}
with $U_\mathrm{LR}$ the Coulomb repulsion energy for electrons in different dots.

Energies in Eqs. \eqref{EQ:11sols0} and \eqref{EQ:02sols0En} are shifted by the gate-voltage controlled energies $\epsilon_\mathrm{L}$ and $\epsilon_\mathrm{R}$. For simplicity we define,
\begin{subequations}
\begin{eqnarray}
\varepsilon &\equiv&  \epsilon_\mathrm{L}-\epsilon_\mathrm{R} -U_\mathrm{RR}+U_\mathrm{LR}, \\
E_{AV}  &\equiv& \frac{1}{2}\left(\epsilon_\mathrm{L}+3\epsilon_\mathrm{R}
+U_\mathrm{LR}+U_\mathrm{RR}\right),
\end{eqnarray}
\end{subequations}
where the detuning, $\varepsilon$, is the difference and $E_{AV}$ is the average between energies $E^{(1,1)}$ and $E^{(0,2)}$ (neglecting the part depending on $n$ and $n'$). Energies for $(0,2)$ and for $(1,1)$ now become:
\begin{subequations}
\bea
E^{(0,2)}_{n,n'}=E_\mathrm{AV}-\ve/2+E_{\mathrm{R}n}+E_{\mathrm{R}n'},\\
E^{(1,1)}_{n,n'}=E_\mathrm{AV}+\ve/2+E_{\mathrm{R}n}+E_{\mathrm{L}n'}.
\label{EQ:11and02detu}
\eea
\end{subequations}
In what follows the common energy shift, $E_\mathrm{AV}$, is omitted because it is irrelevant for the dynamics of the RPE.

In a clean system the single-particle eigenstates in both dots have the same spin and valley properties and therefore the left/right dot part of the (1,1) eigenstates can be separated from the $n$-space part (valley and spin spaces). This separability allows one to write the two-particle eigenstates as products of singlet and triplet functions in spin, valley and dot spaces, see for example the clean case in Ref.~\onlinecite{ReynosoFlensberg2011}. Here, on the other hand, the valley disorder profile and therefore the single particle eigenstates are different in the two dots and each two particle eigenstate---the Slater determinants given above---becomes an arbitrary linear combination of the dot/valley/spin tensor product states. This avoids identifying the eigenstates as singlet and triplet states in spin and valley spaces. However, we will identify singlet-like and triplet-like states according to how they behave when considering single-particle tunneling mixing between $(0,2)$ and $(1,1)$ states. In the next section we see that some particular linear combinations of these Slater determinants---within the subspaces generated by degenerated solutions---are the most physically relevant states.

\section{Interdot tunneling}

\subsection{Disorder induced tunneling between states in different Kramers doublets}

The eigenfunctions of the quantum dot Hamiltonians, $H_0^{\xi}$ (see Eq.\eqref{EQ:Hsingledot}), are $\ketLR{\xi,\mathrm{d},\sigma}$ with $\xi\smeq\{\mathrm{L,R}\}$, $\mathrm{d}\smeq\{+,-\}$, and  $\sigma\!=\{\uparrow,\downarrow\}$. These isolated-dot single-particle solutions become mixed when including tunneling between the dots. In order to proceed, we introduce the identity and Pauli matrices in left/right dot space $\xi_i$, with $i\smeq \{0,1,2,3\}$ (taking $\ketLR{\mathrm{L}}$ and $\ketLR{\mathrm{R}}$ as the $+1$ and $-1$ eigenstates, respectively, of the operator $\xi_3$). Further, for referring to specific blocks of the Hamiltonian we define, $\sigma_\uparrow\equiv{2}^{-1}\left(\sigma_0+\sigma_z\right)$,  $\sigma_\downarrow\equiv{2}^{-1}\left(\sigma_0-\sigma_z\right)$, $\xi_\mathrm{L}\equiv{2}^{-1}\left(\xi_0+\xi_3\right)$, and $\xi_\mathrm{R}\equiv{2}^{-1}\left(\xi_0-\xi_3\right)$.

We assume that the interdot tunneling preserves valley and spin degrees of freedom. This is because the tunneling amplitudes follow from the overlap between the quasi-bounded states in each dot. We take a tunneling energy, $t$, independent of the spin and valley that is being tunneling; this is a good approximation if the height of the confining barrier is much larger than all the energy scales in the DQD Hamiltonian.\cite{WunschDQDCNT2010,WeissFlensberg2010} Then the interdot tunneling Hamiltonian becomes, $H_T=-t \xi_1 \sigma_0 \tau_0$. The complete eight-dimensional single-particle Hamiltonian can now be compactly written as,
\be
H_\mathrm{DQD}=\xi_\mathrm{R} H_0^{\mathrm{R}} +\xi_\mathrm{L} H_0^{\mathrm{L}} -t \xi_1 \sigma_0 \tau_0.
\label{EQ:HDQD}
\ee

In a clean system the isolated-dot solutions of $H_0^{\xi}$, $\ketLR{\xi,\mathrm{d},\sigma}$, have the same spin and valley characteristics in the two dots and tunneling can therefore be described by four independent subsystems of 2 by 2 mixing $\ketLR{\mathrm{L},\mathrm{d},\sigma}$ with $\ketLR{\mathrm{R},\mathrm{d},\sigma}$, which allows one to build the two particle $(0,2)$ and $(1,1)$ eigenstates as tensor products of triplet and singlet functions in left/right dot, valley, and spin spaces.\cite{PalyiB09,PalyiB10,WeissFlensberg2010,WunschDQDCNT2010,ReynosoFlensberg2011} Moreover, the selection rules for the mixing between $(0,2)$ states and $(1,1)$ states follow correspondingly. There is only one avoided crossing produced by the interdot tunneling for each $(0,2)$ state with one $(1,1)$ state.\cite{ReynosoFlensberg2011}

\begin{figure}[t]
\centering
\includegraphics[width=.33\textwidth]{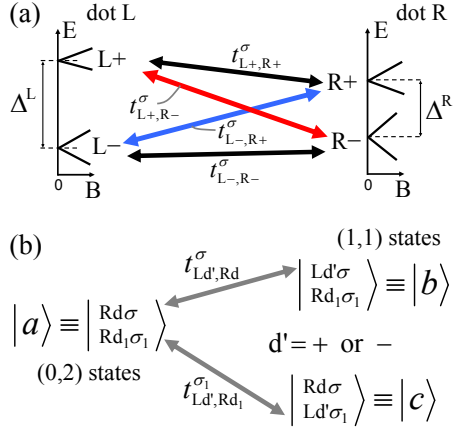}
\caption{(a) For a given disordered sample the valley mixing terms are fixed nonzero values which are different for the two dots: the spin $\sigma$ state in Kramers doublet $\mathrm{L d}$ is a different \emph{valley spinor} than the spin $\sigma$ state in doublet $\mathrm{R d}$. Nonzero tunneling amplitudes appears between $\sigma$ states in different Kramers doublets. (b) For a two-particle $(0,2)$ Slater determinant (which is an eigenstate in the high detuning limit) the interdot tunneling have nonzero matrix elements with four---instead of two---$(1,1)$ Slater determinants.}
\vspace{-0.2cm}
\label{FG:tun}
\end{figure}

In the general disordered case considered here, the disorder in each dot can be different, and therefore the mixing becomes more complicated. We now proceed to the case of zero field (or $ {\bf B}\parallel \zver$). According to Eq.\eqref{EQ:Hsingledot2by2} the Hamiltonian within each dot does not mix opposite spin-$z$ projections, and we have
\be
H_\mathrm{DQD}= \xi_\mathrm{L} \sigma_\uparrow H_\uparrow^\mathrm{L} +\xi_\mathrm{L} \sigma_\downarrow H_\downarrow^\mathrm{L} +\xi_\mathrm{R} \sigma_\uparrow H_\uparrow^\mathrm{R} +\xi_\mathrm{R} \sigma_\downarrow H_\downarrow^\mathrm{R} -t \xi_1 \sigma_0 \tau_0.~~
\ee
Each valley operator $H_\sigma^\xi$ has the valley solutions, $\ketLR{\hat{\varsigma}^{\xi}_{\sigma},\mathrm{d}}$, given in Eq.\eqref{EQ:eigen}. Using as a basis the eigenstates of each dot, we see that the interdot mixing matrix elements $t^{\sigma}_{\mathrm{L d_L},\mathrm{R d_R}} \equiv\braLR{\mathrm{L},\mathrm{d_L},\sigma}H_T\ketLR{\mathrm{R},\mathrm{d_R},\sigma}$ are
\bea
t^{\sigma}_{\mathrm{L d_L},\mathrm{R d_R}} & =&  \left(t^{\sigma}_{\mathrm{R d_R},\mathrm{L d_L}}\right)^*  \label{EQ:tun}\\
&=&-t \braLR{\hat{\varsigma}^\mathrm{L}_{\sigma},\mathrm{d_L}} \ketNR{\hat{\varsigma}^\mathrm{R}_{\sigma},\mathrm{d_R}} \nonumber \\
&=&-t\left(
  \cos\frac{\varphi_{RL}}{2}\cos\left(\frac{\alpha^R_{\mathrm{d}_R,\sigma}-\mathrm{d}_R\mathrm{d}_L\alpha^L_{\mathrm{d}_L,\sigma}}{2}\right)\right.\nonumber\\
&& \left.+\ci \sin\frac{\varphi_{RL}}{2} \cos\left(\frac{\alpha^R_{\mathrm{d}_R,\sigma}+\mathrm{d}_R\mathrm{d}_L\alpha^L_{\mathrm{d}_L,\sigma}}{2}\right) \right),  \nonumber
\eea
where $\varphi_{RL} \smeq \varphi_{\tiny KK'}^{R}\smmi \varphi_{\tiny KK'}^{L}$. As in the clean case, the state $\ketLR{\mathrm{R},\mathrm{d},\sigma}$ can mix with the state $\ketLR{\mathrm{L},\mathrm{d},\sigma}$, but now also with the state $\ketLR{\mathrm{L},\mathrm{\bar{d}},\sigma}$. This situation is sketched in Fig.\ref{FG:tun}(a).

In order to derive properties of the tunneling amplitudes, we arrange the four quantities, $t^{\sigma}_{\mathrm{L d},\mathrm{R d'}}/(-t)$, as a 2 by 2 matrix in doublet space, $D_{\sigma,\mathrm{LR}}$. Using the unitary transformations, $U^{\xi}_{\sigma}$, that diagonalize the valley operators $H^{\xi}_{\sigma}$, one can readily show that $D_{\sigma,\mathrm{LR}}$ is a unitary matrix. The same holds for the matrix $D_{\sigma,\mathrm{RL}}$ ($= D_{\sigma,\mathrm{LR}}^\dagger$). This allow us to write,
\be
\modulo{t^{\sigma}_{L\overline{\mathrm{d}},R\mathrm{d}}}^2+\modulo{\vphantom{t^{\sigma}_{L\overline{\mathrm{d}},R\mathrm{d}}}t^{\sigma}_{L\mathrm{d},R\mathrm{d}}}^2=\modulo{t^{\sigma}_{L\mathrm{d},R\overline{\mathrm{d}}}}^2+\modulo{\vphantom{t^{\sigma}_{L\overline{\mathrm{d}},R\mathrm{d}}}t^{\sigma}_{L\mathrm{d},R\mathrm{d}}}^2=\modulo{t}^2.
\ee
In addition, the orthogonality relation, $\alpha^\gamma_{+,\sigma}\smpl\alpha^\gamma_{-,\sigma}\smeq \pi$,  combined (for zero magnetic field) with the time reversal symmetry relation, $\alpha^\gamma_{\mathrm{d},\sigma}\smeq \alpha^\gamma_{\overline{\mathrm{d}},\overline{\sigma}}$,
lead to the following properties,
\bea
t^{\sigma}_{L\mathrm{d},R\mathrm{d}}= \parente{\vphantom{t^{\sigma}_{L\overline{\mathrm{d}},R\overline{\mathrm{d}}}}
t^{\overline{\sigma}}_{L\mathrm{d},R\mathrm{d}}}^*=\parente{t^{\sigma}_{L\overline{\mathrm{d}},R\overline{\mathrm{d}}}}^*,
\nonumber \\
t^{\sigma}_{L\overline{\mathrm{d}},R\mathrm{d}}= -\parente{\vphantom{t^{\sigma}_{L\overline{\mathrm{d}},R\overline{\mathrm{d}}}} t^{\overline{\sigma}}_{L\overline{\mathrm{d}},R\mathrm{d}}}^*= -\parente{\vphantom{t^{\sigma}_{L\overline{\mathrm{d}},R\overline{\mathrm{d}}}} t^{\sigma}_{L\mathrm{d},R\overline{\mathrm{d}}}}^*.
\eea
In what follows we use these expressions, and their relation to the parameters of the problem, to quantify the effect of the disorder in the return probability experiment.

\subsection{Disorder induced avoided crossings in the two-particle spectrum}

In Eq.\eqref{EQ:tun} we have shown that the disorder induces single-particle interdot tunneling that does not conserve the Kramers doublet index. Before presenting the effect on the mixing between the two particle states, we perform unitary transformations in the doublet basis of each dot in order to work with a real tunneling matrix. This greatly simplifies the notation in the following.

Instead of using as a basis the wavefunctions, $\ketLR{\xi,\mathrm{d},\sigma}$, presented in Eq.\eqref{EQ:eigen}, we multiply each of them by a phase factor,
\be
\ketLR{\xi,\mathrm{d},\sigma}_\mathrm{new} = {\rm e}^{\ci \mu^{\xi}_{\mathrm{d},\sigma} } \ketLR{\xi,\mathrm{d},\sigma}_\mathrm{old},
\label{EQ:newEig}
\ee
and the new matrix elements become,
\be
\left(t^{\sigma}_{L\mathrm{d},R\mathrm{d'}}\right)_\mathrm{new}=\left(t^{\sigma}_{L\mathrm{d},R\mathrm{d'}}\right)_\mathrm{old} {\rm e}^{\ci \left(\mu^{\mathrm{R}}_{\mathrm{d'},\sigma}-\mu^{\mathrm{L}}_{\mathrm{d},\sigma} \right) }.
\ee
There are infinite choices for making the new tunneling amplitudes real numbers. We choose,
\bese
\bea
\mu^{\mathrm{R}}_{\mathrm{-},\sigma}&=&0, \\
\mu^{\mathrm{L}}_{\mathrm{-},\sigma}&=&\mathrm{Arg}\left[\left(-t^{\sigma}_{L\mathrm{-},R\mathrm{-}}\right)_\mathrm{old}\right] ,\\
\mu^{\mathrm{L}}_{\mathrm{+},\sigma}&=&\mathrm{Arg}\left[\left(-t^{\sigma}_{L\mathrm{+},R\mathrm{-}}\right)_\mathrm{old}\right] ,\\
\mu^{\mathrm{R}}_{\mathrm{+},\sigma}&=&\mu^{\mathrm{L}}_{\mathrm{-},\sigma}+\mu^{\mathrm{L}}_{\mathrm{+},\sigma}, \eea
\eese
and then we have,
\bese
\bea
\left(t^{\sigma}_{\mathrm{L d},\mathrm{R d}}\right)_\mathrm{new}=-t \cos\frac{\eta_\sigma}{2},\\
\left(t^{\sigma}_{\mathrm{L d},\mathrm{R \bar{d}}}\right)_\mathrm{new}=-t \mathrm{d} \sin\frac{\eta_\sigma}{2},
\eea
\label{EQ:NewTunAmpls}
\eese
with $\eta_\sigma$ the angle between the vectors in valley space: $\hat{\varsigma}^\mathrm{R}_{\sigma}$ and $\hat{\varsigma}^\mathrm{L}_{\sigma}$.
The amplitudes $\left(t^{\sigma}_{\mathrm{R d},\mathrm{L d'}}\right)_\mathrm{new}$ follow straightforwardly  by conjugating the matrix elements above.

A physical intuitive (and equivalent) approach for writing the tunneling amplitudes as real numbers follows from: (i) perform a unitary transformation \emph{in both dots} making the $z$-axis coincide with one of the valley vectors, for example $\hat{\varsigma}^\mathrm{R}_{\sigma}$; (ii) rotate both dots around the new $z$-axis in order to take the other valley vector, i.e., $\hat{\varsigma}^\mathrm{L}_{\sigma}$, to the $xz$-plane where its eigenvectors can be written as real valley spinors fully described by the angle, $\eta_\sigma$, between the vectors $\hat{\varsigma}^\mathrm{\xi}_{\sigma}$; (iii) leave the valley basis fixed in the first dot ($\xi\smeq\mathrm{R}$) and perform a unitary transformation in the other dot ($\xi\smeq\mathrm{L}$) making the real eigenvectors the new basis; (iv) the resulting tunneling amplitudes become real.

\begin{figure*}[!ht]
   \centering
   \includegraphics[width=.84\textwidth]{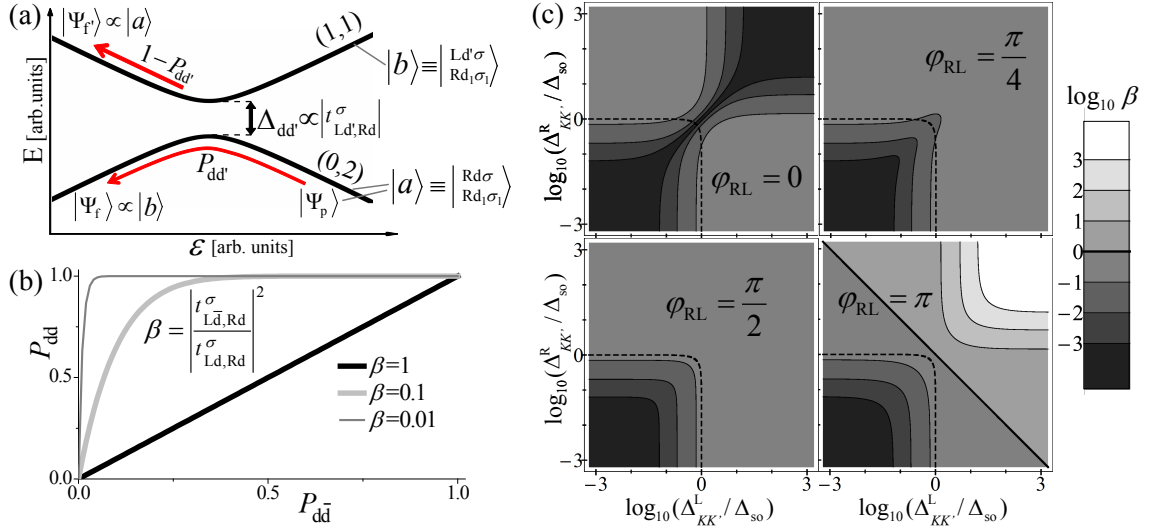}
   \vspace{-0.3cm}
  \caption{(a) Simplest case for an avoided crossing as a result of the mixing between a $(0,2)$ state ($\ketLR{a}$) and a $(1,1)$ state ($\ketLR{b}$). A detuning pulse takes the system, initialized in $\ketLR{a}$, from positive to negative detuning. The Landau-Zener probability to follow the avoided crossing is, $P_{\mathrm{d d'}}$. (b) Probability, $P_{\mathrm{d d}}$ (in a DCAC) as a function of $P_{\mathrm{d \bar{d}}}$ (in a disorder induced avoided crossing, i.e., a DFAC), for different ratios between the gaps, $\beta\equiv \left(\Delta_{\mathrm{d \bar{d}}}/\Delta_{\mathrm{d d}}\right)^2$, assuming that both processes are realized at the same detuning speed. One sees that $\beta$, which quantifies the disorder, must be very small for the probabilities $P_{\mathrm{d \bar{d}}}$ to be neglected and while simultaneously assuming that $P_{\mathrm{d d}}$ is close to 1 (adiabatic process). (c) Color plot of $\beta$ as a function of the parameters of the problem for ${\bf B}\smeq 0$. We see that moderated values of the valley mixing parameters can generate non-negligible values of $\beta$ and therefore the LZ physics in the disorder induced avoided crossings must be in general taken into account.}
   \label{FG:Mfac1}
   \vspace{-0.2cm}
\end{figure*}

From hereon we use the \emph{new} single particle eigenstates of the isolated dots presented in Eq.\eqref{EQ:newEig} for the derivations involving two particle states.
\subsection{Landau-Zener physics for different kind of crossings}
We now apply the tunneling Hamiltonian, its two particle version [$H_{T}^{\rm 2p}=\eins_{\mathbf{1}}\otimes ({H_{T}^{\rm 1p}})\vphantom{\eins}_{\mathbf{2}}+({H_{T}^{\rm 1p}})\vphantom{\eins}_{\mathbf{1}}\otimes \eins_{\mathbf{2}}$], to the $(0,2)$ eigenstates,
\be
H_{T}^{\rm 2p} \ketLR{{}_{\mathrm{R,d_1},\sigma_1}^{\mathrm{R,d},\sigma}} =  \sum_{\mathrm{d'}=\pm}{\left[t^{\sigma_1}_{\mathrm{L d'},\mathrm{R d_1}}\ketLR{{}_{\mathrm{L,d'},\sigma_1}^{\mathrm{R,d},\sigma}}+ t^{\sigma}_{\mathrm{L d'},\mathrm{R d}}\ketLR{{}_{\mathrm{R,d_1},\sigma_1}^{\mathrm{L,d'},\sigma}}\right]}.
\label{Eq:TunAction}
\ee
This result is sketched in Fig.\ref{FG:tun}(b) where we show that, by virtue of the disorder, each $(0,2)$ Slater determinant have nonzero matrix elements with four , instead of two, $(1,1)$ Slater determinants.

We now focus on the disorder induced Landau-Zener processes that arise due to the additional mixing terms. For simplicity, in this first approximation to the problem, we assume that the energies, $E_l$, of the four $(1,1)$ Slater determinants (we label them $l\smeq\{1,2,3,4\}$) are well separated on the scale of the tunneling $t$ and therefore we can write four separated effective Hamiltonians describing the mixing with the $(0,2)$ state as,
\begin{equation}
H_{l}=\left(\begin{array}{cc}-\frac{\varepsilon}{2}
&-t_l\\
-t_l&\frac{\varepsilon}{2}+E_l
\end{array}\right).
\label{EQ:H2by2}
\end{equation}
At the crossing energies the mixing due to interdot tunneling is maximum and avoided crossings with gaps $\Delta_l= 2|t_l|$ appear. By inspecting Eqs.\eqref{EQ:NewTunAmpls} and \eqref{Eq:TunAction} we readily identify two different avoided crossings types; (a) the doublet conserving avoided crossings (DCACs), with a gap value of, $\Delta^{\sigma}_{\mathrm{dd}}\smeq 2|t\cos({\eta_\sigma}/{2})|$; and (b) the doublet-flipping avoided crossings (DFACs), with a gap value of, $\Delta^{\sigma}_{\mathrm{d\bar{d}}}\smeq 2|t\sin({\eta_\sigma}/{2})|$. The latter type, DFACs, is the one induced by disorder because when the two valley vectors are the same, as in a clean case, we have $\eta_\sigma\smeq 0$ and $\Delta_{\mathrm{d\bar{d}}}\smeq0$.

As described in the Introduction and in Fig.\ref{FG::f1}, detuning changes are intrinsic to the RPE single-shot cycle: the separation and joining stages. The gap is an important parameter for determining the probability of realizing an state conversion when a change of detuning is applied and the system passes through an avoided crossing. In a given stage of the problem, $h$ (with $h=\mathrm{s,j}$, \emph{separation} and \emph{joining} stages, respectively), the Landau-Zener probability of \emph{realizing} the state conversion in an avoided crossing, $l$, is,
\bese
\bea
P^{(h)}&=&1-\overline{P}^{(h)},\\
\overline{P}^{(h)}&\equiv&\exp{\left[-\frac{2\pi \Delta_l^2}{\hbar v^{(h)}_l}\right]}.
\eea
\label{EQ:LZformula}
\eese
The probability $P^{(h)}$ grows the slower is the rate of change of detuning at the crossing $l$, $v^{(h)}_l$, and the bigger is the gap of the avoided crossing, $\Delta_l$. For convenience, we have defined $\overline{P}^{(h)}$ as the probability to remain in the original state, this probability approaches zero the more adiabatically the Landau-Zener process is realized.

Because the gaps in the exponents of the Landau-Zener formula appear to the second power we define,
\be
 \beta^\sigma\equiv \left(\frac {\Delta^\sigma_{\mathrm{d\bar{d}}}}{\Delta^{\sigma}_{\mathrm{d d}}}\right)^2\smeq \tan^2\left(\eta_\sigma/2\right),
\ee
This number allows us to quantify the effect of the valley mixing for the Landau-Zener physics involved in the experiment. How sensitive $\beta^\sigma$ is with the parameters in the double dot is an important question. Since the angle between the valley fields is given by,
\be
\eta_\sigma = \arccos\left( \frac{\Delta^\mathrm{L}_{\tiny KK'} \Delta^\mathrm{R}_{\tiny KK'} \cos \varphi_\mathrm{RL} +{\delta}_{\sigma}^\mathrm{R} {\delta}_{\sigma}^\mathrm{L}}{\Delta^\mathrm{L}_{\sigma} \Delta^\mathrm{R}_{\sigma}}\right);
\label{EQ:etaAngl}
\ee
by virtue of the half angle relations, we find that $\beta^\sigma$ is given by the quotient,
\be
\beta^{\sigma}=\frac{\Delta^\mathrm{L}_{\sigma} \Delta^\mathrm{R}_{\sigma}-\Delta^\mathrm{L}_{\tiny KK'} \Delta^\mathrm{R}_{\tiny KK'} \cos \varphi_\mathrm{RL} - \delta_{\sigma}^\mathrm{L}\delta_{\sigma}^\mathrm{R}}{\Delta^\mathrm{L}_{\sigma} \Delta^\mathrm{R}_{\sigma}+\Delta^\mathrm{L}_{\tiny KK'} \Delta^\mathrm{R}_{\tiny KK'} \cos \varphi_\mathrm{RL} + \delta_{\sigma}^\mathrm{L}\delta_{\sigma}^\mathrm{R}}.
\label{EQ:betaFactor}
\ee

When ${\bf B}\smeq 0$, there is no dependence with $\sigma$ in the angle $\eta_\sigma$ and consequently, nor in the tunnelings amplitudes and the gaps. In what follows, we assume that the lengths of the two dots are the same and thus the spin-orbit splittings for the lowest lying bound states are identical in both dots; i.e., $\delta_{\sigma}^\mathrm{R}\smeq\delta_{\sigma}^\mathrm{L}=\bar{\sigma}\Delta_{so}$. However, it must be noted that if the dots have different lengths leading to different spin-orbit splittings, then the effect of disorder in the Landau-Zener physics can be even more drastic than what we show below, because the angle $\eta_\sigma$ becomes more sensitive to the in-plane valley fields differences. Our formulas in Eqs.\eqref{EQ:betaFactor} and \eqref{EQ:etaAngl} are general and can be used for such cases.

Figure \ref{FG:Mfac1}(a) shows a sketch of an avoided crossing between a $(0,2)$ state and a $(1,1)$ state reflecting the Hamiltonian presented in Eq.\eqref{EQ:H2by2}. In the example, the avoided crossing mixes a $(0,2)$ state, $\ketLR{a}$, and with an state $(1,1)$, $\ketLR{b}$. Note that the $\ketLR{b}$ Slater determinant differs from the $\ketLR{a}$ one in that a $\sigma$ single particle right-dot eigenstate with double index $\mathrm{d}$ is replaced by a $\sigma$ left-dot eigenstate with the doublet index $\mathrm{d'}$; thus the mixing is due to the $t^{\sigma}_\mathrm{L d',R d}$  tunneling amplitude. The system is prepared, at positive detuning, in the state $\ketLR{\Psi_\mathrm{p}}\smeq \ketLR{a}$ and then the detuning is swept to negative values. The Landau-Zener process governs the probabilities that the final measured state, $\ketLR{\Psi_\mathrm{f}}$, reflects a state conversion, $\ketLR{\Psi_\mathrm{f}}\propto\ketLR{b}$ (with $P_\mathrm{dd'}$ as given in Eq.\eqref{EQ:LZformula}), or it remains in the original state,  $\ketLR{\Psi_\mathrm{f}}\propto\ketLR{a}$ (with $\overline{P}_\mathrm{dd'}\smequiv 1-{P}_\mathrm{dd'}$).

Figure \ref{FG:Mfac1}(b) shows the Landau-Zener probabilities (see Eq.\eqref{EQ:LZformula}) in a DCAC versus the probability in a DFAC linked by different values of $\beta$ assuming both Landau-Zener processes were performed at the same detuning rate. The result shows that only for $\beta\ll 0.01$ one could ignore the disorder induced avoided crossings assuming that $P_\mathrm{d\bar{d}}\approx 0$ and at the same time assume that the doublet conserving crossings provide adiabatic state conversions, i.e., $P_\mathrm{dd}\approx 1$. In panel (c) we present a map of $\beta_{\sigma}$ $(=\!\beta_{\bar{\sigma}}=\!\beta)$, from Eq.\eqref{EQ:betaFactor}, for different phase differences between the valley mixing terms and as a function of their strengths, $\Delta_{\tiny KK'}^{\xi}$, normalized to the spin-orbit coupling splitting. The plot allows to see that $\beta$ can be above 0.01---i.e., the Landau-Zener physics in the DFACs cannot be neglected---in a broad region in parameter space, even in situations in which the spin-orbit coupling dominates in both quantum dots.

\subsection{Full spectrum, classification of the avoided crossings}
\label{SC:classification}
\begin{figure}[t]
   \centering
   \includegraphics[width=.38\textwidth]{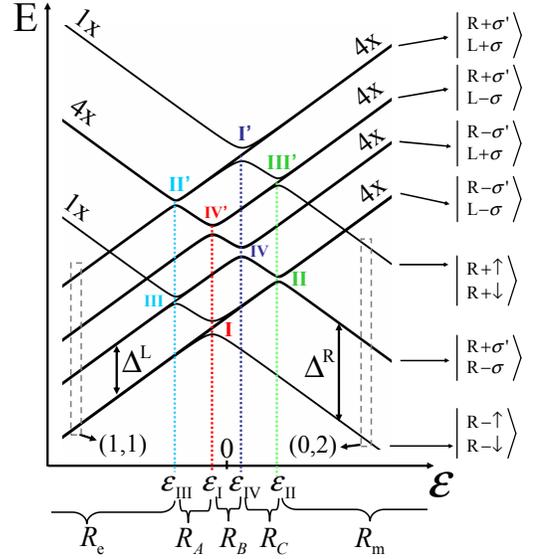}
   \vspace{-0.4cm}
      \caption{Avoided crossings classification in the two particle spectrum for zero magnetic field reflecting the mixing between the six $(0,2)$ states and the sixteen $(1,1)$ states as a function of detuning. The avoided crossings III, III', II and II' are induced by disorder, they rely on tunneling matrix elements that do not conserve doublet index. In the example $\Delta^\mathrm{L}/\Delta^\mathrm{R}\smeq 0.55$ and $\beta\smeq 0.29$.}
   \vspace{-0.2cm}
  \label{FG:detun}
\end{figure}
As shown above for the zero field case, the Landau-Zener physics in doublet-flipping avoided crossings can be important for the return probability experiment and therefore their effects must be studied. If the tunneling energy, $t$, is larger than the splitting energies in the dots, $\Delta^\xi$, the mixing of each $(0,2)$ states with the four $(1,1)$ states given in Eq.\eqref{Eq:TunAction} must be considered simultaneously. No simple Landau-Zener physics concepts can be applied because the mixing cannot be decomposed into two-level avoided crossings. In what follows, we study the problem assuming that $t$ is smaller than the $\Delta^\xi$ energies and that the valley mixing is sufficiently different in the two quantum dots so that $|\Delta^\mathrm{L}-\Delta^\mathrm{R}|>t$. In this limit, the 22 levels two-particle spectrum [describing the mixing of the six $(0,2)$ states with the sixteen $(1,1)$ states] can be decomposed into two-level avoided crossings. The LZ formula can be used to describe all the processes in the different avoided crossings happening as the separation and the joining stages of the RPE are performed. The qualitative conclusions drawn for the low $t$ picture can be extended to the general $t$ case.  We also note that for zero magnetic field the tunneling amplitudes given in Eq.\eqref{EQ:NewTunAmpls} can be chosen to be independent of $\sigma$, therefore, in what follows we omit the superindex $\sigma$ and set $t_\mathrm{L d',R d} \equiv t^\sigma_\mathrm{L d',R d}$.

The small $t$ assumption allows us to introduce a classification of the avoided crossings. In Fig.\ref{FG:detun} we show such a classification in the two particle spectrum as a function of detuning, obtained by taking the two particle version of the Hamiltonian of Eq.\eqref{EQ:HDQD} and using the Slater determinants constructed from the valley and spin single particle states,  $\ketLR{\xi\tau\sigma}\smeq \ketLR{\xi}\otimes\ketLR{\tau}\otimes\ketLR{\sigma}$, i.e., the six independent $(0,2)$ states $\ketLR{{}^{\mathrm{R}\tau\sigma}_{\mathrm{R}\tau'\sigma'}}$ with $(\tau'\sigma')\smneq(\tau\sigma)$, plus the sixteen $(1,1)$ Slater determinants,
$\ketLR{{}^{\mathrm{L}\tau\sigma}_{\mathrm{R}\tau'\sigma'}}$. The avoided crossings presented in the figure are studied in detail below, their associated detuning and energy values are given in Table \ref{TB:charC}.

From Figure \ref{FG:detun} it is clear that in the high detuning limit the sixteen $(1,1)$ states are grouped into four subsets of four states given by the doublets that are occupied in the left and the right dot, $\mathrm{d_L}$ and $\mathrm{d_R}$, respectively. We denote those four subsets as, $(\mathrm{L d_L,R d_R})$, with states, $\ketLR{{}^{\mathrm{L d_L}\sigma}_{\mathrm{R d_R}\sigma'}}$, and energies,
\be
E^{(1,1)}_\mathrm{L d_L,R d_R}\smeq \varepsilon/2+ \mathrm{d_L}\Delta^\mathrm{L}/2+\mathrm{d_R}\Delta^\mathrm{R}/2.
\label{EQ:E11}
\ee
Similarly, the energies of the $(0,2)$ states in the high detuning limit are,
\be
E^{(0,2)}_\mathrm{R d_1,R d_2}\smeq -\varepsilon/2+ \mathrm{d_1}\Delta^\mathrm{R}/2+\mathrm{d_2}\Delta^\mathrm{R}/2.
\label{EQ:E02}
\ee
For $\mathrm{d_1}\smeq -\mathrm{d_2}$ the energy level, $-\varepsilon/2$, is fourfold degenerated since it is associated with the four states, $\ketLR{{}^{\mathrm{R}+\sigma}_{\mathrm{R}-\sigma'}}$. For future reference we denote this subset as $(\mathrm{R +,R -})$.

We first focus on the avoided crossings involving the two $(0,2)$ non-degenerated states (with high detuning energies in Eq.\eqref{EQ:E02} taking $\mathrm{d_1}\smeq\mathrm{d_2}\smeq\pm$), namely, the ground state and the highest excited $(0,2)$ states,
\be
\ketLR{{S}^{(0,2)}_\mathrm{d d} } \equiv \ketLR{{}_{\mathrm{R d }\downarrow}^{\mathrm{R d} \uparrow}}.
\label{EQ:singlets02}
\ee
Because the two electrons occupy the same Kramers doublet $\mathrm{d}$ the total spin is zero. We use the letter $S$ for these singlet-like solutions even though these states \emph{are not} singlets in spin space: the spin part of the two particle state cannot be separated even in the absence of valley mixing.\cite{PalyiB09,ReynosoFlensberg2011}
Following Eq.\eqref{Eq:TunAction} and applying the low $t$ picture, we can treat separately the mixing due to the tunneling Hamiltonian between the states $\ketLR{{S}^{(0,2)}_\mathrm{d d} }$ and $(1,1)$ states in different levels $(\mathrm{L d_L,R d_R})$. Since the single particle tunneling is unable to change two single particle states simultaneously, the state $\ketLR{{S}^{(0,2)}_\mathrm{d d} }$ does not mix with the states in subsets $(\mathrm{L d_L,R \bar{d}})$.

Each doublet conserving tunneling amplitude,
$t_\mathrm{L d, R d}$, mixes the state $\ketLR{{S}^{(0,2)}_\mathrm{d d} }$ with the $(1,1)$ combination,
\be
\ketLR{{S}^{(1,1)}_\mathrm{d d} } \equiv \frac{1}{\sqrt{2}}\left(\ketLR{{}_{\mathrm{L d }\downarrow}^{\mathrm{R d} \uparrow}}+\ketLR{{}_{\mathrm{R d }\downarrow}^{\mathrm{L d} \uparrow}}\right),
\label{EQ:singlet11a}
\ee
of the subset $(\mathrm{L d,R d})$. This defines the I ($\mathrm{d}\smeq-$) and I' ($\mathrm{d}\smeq+$) doublet-conserving avoided crossings shown in Figure \ref{FG:detun}. In this case the matrix element of the tunneling operator is $\sqrt{2} t_\mathrm{L d, R d}$ and therefore the gaps are $\Delta_\mathrm{I}=\Delta_\mathrm{I'}=2 \sqrt{2}\left| \cos \frac{\eta}{2}\right|$.

 Similarly, each tunneling amplitude, $t_\mathrm{L \bar{d}, R d}$, mixes the state $\ketLR{{S}^{(0,2)}_\mathrm{d d} }$ with the $(1,1)$ combination,
\be
\ketLR{{S}^{(1,1)}_\mathrm{\bar{d} d} } \equiv \frac{1}{\sqrt{2}}\left(\ketLR{{}_{\mathrm{L \bar{d} }\downarrow}^{\mathrm{R d} \uparrow}}+\ketLR{{}_{\mathrm{R d }\downarrow}^{\mathrm{L \bar{d}} \uparrow}}\right),
\label{EQ:singlet11b}
\ee
of the subset $(\mathrm{L \bar{d},R d})$. This mixing generates the two doublet-flipping avoided crossings, III ($\mathrm{d}\smeq-$) and III' ($\mathrm{d}\smeq+$), with gaps $\Delta_\mathrm{III}=\Delta_\mathrm{III'}=2 \sqrt{2} \left|\sin \frac{\eta}{2}\right|$.

\begin{table}[b]
\vspace{-0.5cm}
\caption{\label{TB:charC} Detuning and energy values for the avoided crossings presented in Figs.\ref{FG:detun} and \ref{FG:cross}. The detuning position of the avoided crossings are obtained by equating the high detuning energies of Eq.\eqref{EQ:E11} and \eqref{EQ:E02}. The Kramers doublet index conservation (or not) of the tunneling that generates the mixing is also stated.} \centering
\begin{tabular}{ccccc}%{c||l|l|l|l|l|l|l|l}
$\mathrm{C_1}$,~$\mathrm{C_2}$&$\varepsilon_{\mathrm{C_1}}\smeq\varepsilon_{\mathrm{C_2}}$&$E_{\mathrm{C_1}}$&$E_{\mathrm{C_2}}$&Type\\
\hline
\hline
$\mathrm{III}$,~$\mathrm{II'}$&$~-\frac{1}{2}\left(\Delta^\mathrm{L} +\Delta^\mathrm{R}\right)~$&$~\frac{1}{4}\left(\Delta^\mathrm{L}-3\Delta^\mathrm{R}\right)~$&$~\frac{1}{4}\left(\Delta^\mathrm{L} +\Delta^\mathrm{R}\right)~~$&DFAC\\
$\mathrm{III'}$,~$\mathrm{II}$&$~\frac{1}{2}\left(\Delta^\mathrm{L} +\Delta^\mathrm{R}\right)~$&$~\frac{1}{4}\left(3\Delta^\mathrm{R}-\Delta^\mathrm{L}\right)~$&$~-\frac{1}{4}\left(\Delta^\mathrm{L} +\Delta^\mathrm{R}\right)~~$&DFAC\\
$\mathrm{I}$,~$\mathrm{IV'}$&$~\frac{1}{2}\left(\Delta^\mathrm{L} -\Delta^\mathrm{R}\right)~$&$~-\frac{1}{4}\left(\Delta^\mathrm{L}+3\Delta^\mathrm{R}\right)~$&$~\frac{1}{4}\left(\Delta^\mathrm{R} -\Delta^\mathrm{L}\right)~~$&DCAC\\
$\mathrm{I'}$,~$\mathrm{IV}$&$~\frac{1}{2}\left(\Delta^\mathrm{R} -\Delta^\mathrm{L}\right)~$&$~\frac{1}{4}\left(\Delta^\mathrm{L}+3\Delta^\mathrm{R}\right)~$&$~\frac{1}{4}\left(\Delta^\mathrm{L} -\Delta^\mathrm{R}\right)~~$&DCAC\\
 $\mathrm{V}$&0&0&-&DCAC
\end{tabular}
\end{table}

Apart from the factor $\sqrt{2}$ in the gaps, the avoided crossings I, I', III and III' share another important feature. Because in each crossing the single $(0,2)$ state, $\ketLR{{S}^{(0,2)}_\mathrm{d d} }$, crosses four degenerated $(1,1)$ states (and mixes with only one combination of them) three other $(1,1)$ states of the $(\mathrm{L d',Rd})$ subset are unaffected by the interdot tunneling. These states are,
\bese
\bea
\ketLR{{T}^{(1,1)}_{+1,\mathrm{d' d}}} &\equiv& \ketLR{{}_{\mathrm{R d }\uparrow}^{\mathrm{L d'} \uparrow}}, \\
\ketLR{{T}^{(1,1)}_{-1,\mathrm{d' d}}} &\equiv& \ketLR{{}_{\mathrm{R d }\downarrow}^{\mathrm{L d'} \downarrow}}, \\
\ketLR{{T}^{(1,1)}_{0,\mathrm{d' d}}} &\equiv& \frac{1}{\sqrt{2}}\left(\ketLR{{}_{\mathrm{L d' }\downarrow}^{\mathrm{R d} \uparrow}}-\ketLR{{}_{\mathrm{R d }\downarrow}^{\mathrm{L d'} \uparrow}}\right),
\eea
\label{EQ:triplets11}
\eese
with energies growing linear with detuning following Eq.\eqref{EQ:E11}. We denote them with $T$ in direct analogy with the spin-triplet states of the $(1,1)$ configuration in spin-only double dots; in such a situation the interdot tunneling only affects the spin-singlets $(0,2)$ and $(1,1)$ leaving the spin triplet states unaffected. In the language of Pauli-blockade these are blocked states.\cite{Johnson2005,Koppens2005,PalyiB09} We will refer to them as triplet-like states despite the fact that $\ketLR{{T}^{(1,1)}_\mathrm{0,d' d}}$ is not a spin-triplet.

\begin{figure}[t]
   \centering
   \includegraphics[width=.38\textwidth]{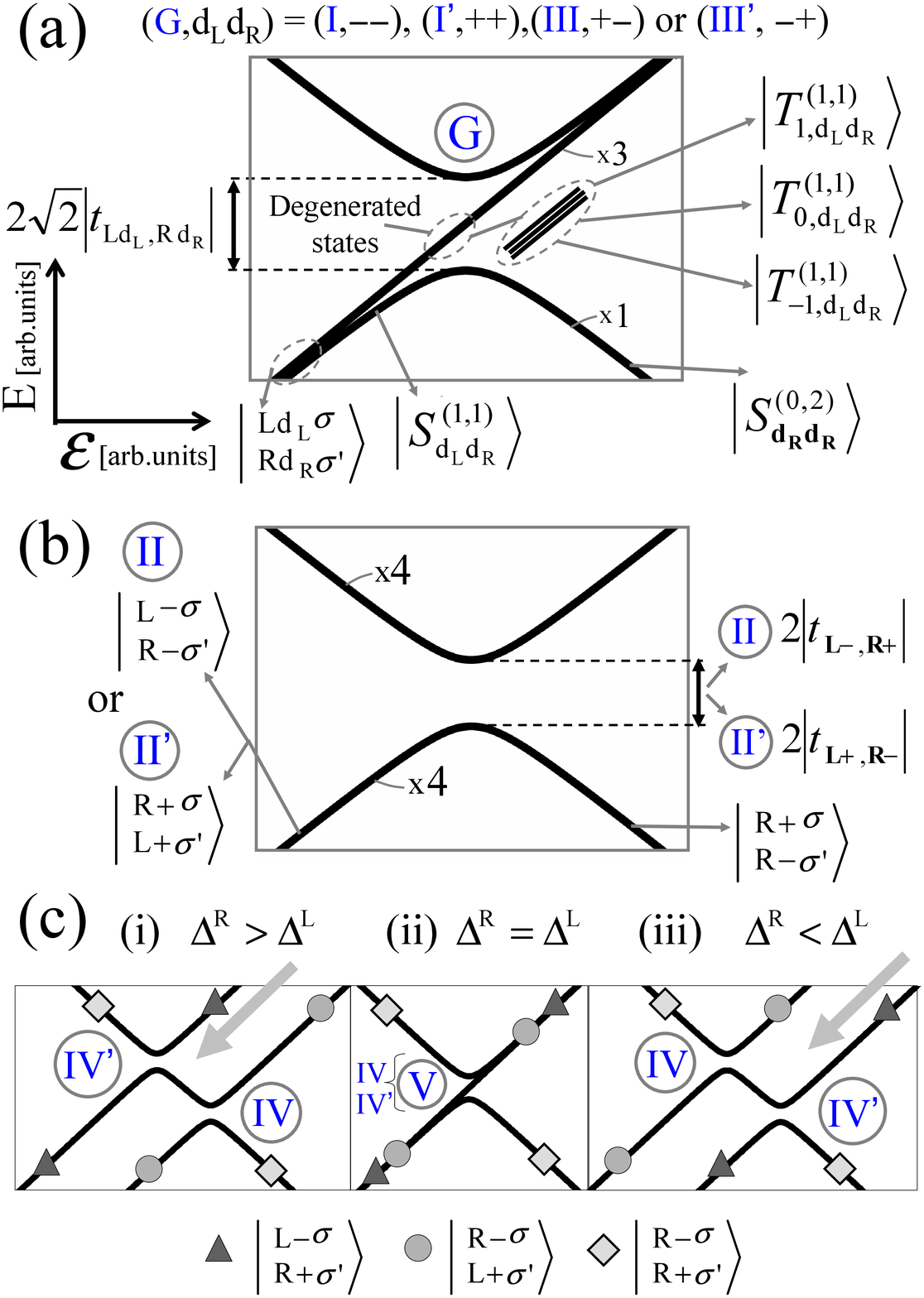}
   \vspace{-0.3cm}
   \caption{States involved in the avoided crossings presented in Fig.\ref{FG:detun}. (a) Avoided crossings $\mathrm{G= I,I',III~or~III'}$, the last two are induced by disorder. As a single $(0,2)$ state crosses four $(1,1)$ states there are three $(1,1)$ blocked states---not mixed by interdot tunneling---in close analogy with spin-only double dots. (b) No blocked states appear in the disorder-induced avoided crossings II and II', four $(0,2)$ states cross four $(1,1)$ states. (c) The doublet conserving avoided crossings IV and IV' also do not present blocked states. According to the values $\Delta^\mathrm{L}$ and $\Delta^\mathrm{R}$ the detuning dependence changes which lead to different outcomes in the RPE; we refer to the gray arrows in Sec.\ref{SC:ResPrepHS} when showing the two possible situations available if $\ketLR{{S}^{(0,2)}_{++}}$ is prepared. When $\Delta^\mathrm{R}\smeq\Delta^\mathrm{L}$ (extremely unlike in a  disordered system) the $(\mathrm{L +,R -})$ and $(\mathrm{L -,R +})$ subsets become degenerated and IV and IV' would collapse in a single crossing, V.}
   \vspace{-0.2cm}
   \label{FG:cross}
\end{figure}

We now focus on the crossings involving the remaining $(0,2)$ states, i.e., states of the $(\mathrm{R +,R -})$ subset. Again, the small $t$ picture allows us to study separately the four crossings with the $(\mathrm{L d_L,R d_R})$ subsets of the $(1,1)$ configuration. By inspecting the action of the interdot tunneling operator on a $(0,2)$ eigenstate given in Eq.\eqref{Eq:TunAction} one can show that all the crossings become avoided crossings and no blocked states remain. This is easily seen by noting that any $(1,1)$ state $\ketLR{{}_{\mathrm{L d_L},\sigma'}^{\mathrm{R d_R}\sigma}}$ is mixed in an effective 2 by 2 system---due to interdot tunneling amplitude, $t_{\mathrm{L d_L},\mathrm{R \bar{d}_R}}$---with the $(0,2)$ state $\ketLR{{}_{\mathrm{R \bar{d}_R},\sigma'}^{\mathrm{R d_R}\sigma}}$. The singlet-like and triplet-like basis defined in Eqs.\eqref{EQ:singlet11a}, \eqref{EQ:singlet11b}, \eqref{EQ:triplets11} facilitates the calculation of the hyperfine dynamics. For the avoided crossings here, we have
\bese
\bea
\braLR{{T}^{(1,1)}_{\sigma_t,\mathrm{d_L d_R}}} H_{T}^{\rm 2p} \ketLR{{T}^{(0,2)}_{\sigma_t,\mathrm{\bar{d}_R d_R}}}&=&t_{\mathrm{L d_L},\mathrm{R \bar{d}_R}}, \\
\braLR{{S}^{(1,1)}_{\mathrm{d_L d_R}}} H_{T}^{\rm 2p} \ketLR{{S}^{(0,2)}_{\mathrm{\bar{d}_R d_R}}}&=&t_{\mathrm{L d_L},\mathrm{R \bar{d}_R}},
\eea
\label{EQ:HT2pmix}
\eese
with $\sigma_t=\{-1,0,1\}$ and,
\bea
\ketLR{{T}^{(0,2)}_{+1,\mathrm{\bar{d} d}}} \equiv\ketLR{{}_{\mathrm{R d }\uparrow}^{\mathrm{R \bar{d}} \uparrow}}~&,&~~
\ketLR{{T}^{(0,2)}_{-1,\mathrm{\bar{d} d}}} \equiv \ketLR{{}_{\mathrm{R d }\downarrow}^{\mathrm{R \bar{d}} \downarrow}},
\label{EQ:singANDtrip02}
\\
\ketLR{{T}^{(0,2)}_{0,\mathrm{\bar{d} d}}} \equiv \frac{1}{\sqrt{2}}\left(\ketLR{{}_{\mathrm{R \bar{d} }\downarrow}^{\mathrm{R d} \uparrow}}-\ketLR{{}_{\mathrm{R d }\downarrow}^{\mathrm{R \bar{d}} \uparrow}}\right)&,& ~~\ketLR{{S}^{(0,2)}_{\mathrm{d \bar{d}}}} \equiv \frac{1}{\sqrt{2}}\left(\ketLR{{}_{\mathrm{R \bar{d} }\downarrow}^{\mathrm{R d} \uparrow}}+\ketLR{{}_{\mathrm{R d }\downarrow}^{\mathrm{R \bar{d}} \uparrow}}\right),~ \nonumber
\eea
i.e., we have defined triplet-like and singlet-like states inside the fourfold degenerated subset $(\mathrm{R +,R -})$.

In particular, the doublet conserving tunneling amplitudes generate avoiding crossings with $(1,1)$ states in the $(\mathrm{L +,R -})$ and $(\mathrm{L -,R +})$ subsets; in Fig.\ref{FG:detun} we have label these avoided crossings as IV and IV', respectively. Their gaps are given by $\Delta_\mathrm{IV}=\Delta_\mathrm{IV'}=2 \left|t\cos \frac{\eta}{2}\right|$.
Conversely, the doublet-flipping avoided crossings II and II' have gaps, $\Delta_\mathrm{II}=\Delta_\mathrm{II'}=2 \left|t\sin \frac{\eta}{2}\right|$,
and are associated with the crossings between $(0,2)$ states belonging to the $(\mathrm{R +,R -})$ subset and $(1,1)$ states belonging to the subsets $(\mathrm{L -,R -})$ and $(\mathrm{L +,R +})$, respectively.

In Fig.\ref{FG:cross} we present a detailed sketch of all the mentioned crossings that include the relevant states. Figure \ref{FG:cross}(c) reflects the fact that, as it is shown in Table \ref{TB:charC}, the relative positions of crossings IV and IV' depends on the difference between $\Delta^\mathrm{R}$ and $\Delta^\mathrm{L}$. This happens because the avoided crossing involving the (0,2) subset $(\mathrm{R +,R -})$ is IV when crossing $(\mathrm{L +,R -})$ and IV' when crossing $(\mathrm{L -,R +})$. As it is shown in the following, this leads to different outcomes of the return probability experiment if excited (0,2) states are prepared. Finally, in the central panel of Fig.\ref{FG:cross}(c) we show that IV and IV' collapse into a single avoided crossing, V, when  $\Delta^\mathrm{R}\smeq\Delta^\mathrm{L}$. We do not consider this avoided crossing further because such a special case is not representative for a disordered system.
\begin{figure*}[t]
   \centering
   \includegraphics[width=.94\textwidth]{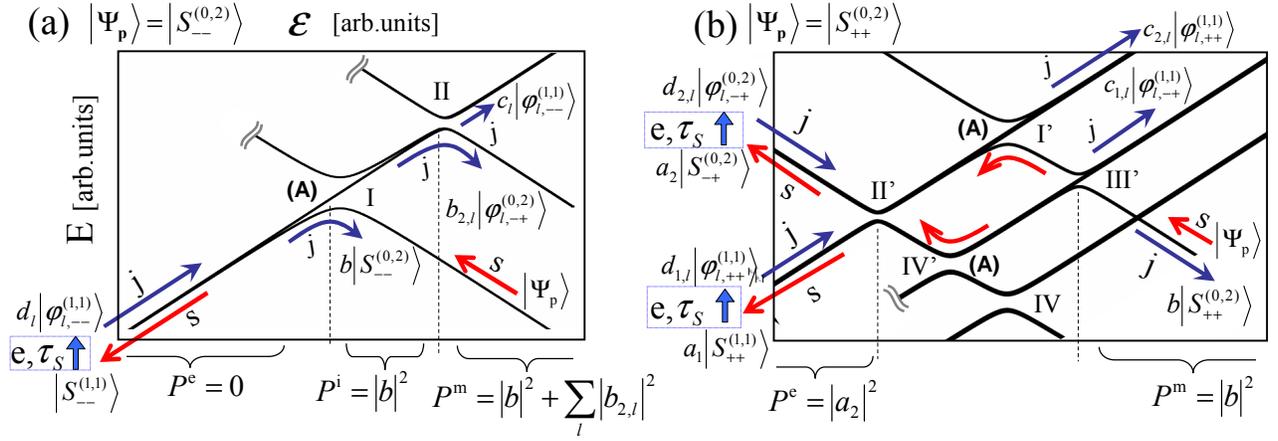}
   \vspace{-0.4cm}
   \caption{Single-shot return probability experiment: separation (``s"), evolution (``e", during a time $\tau_s$) and joining (``j") stages. Avoiding crossings (A) are doublet-conserving and, in the situation we study, their gaps are much bigger than the others; the state conversions in crossings (A) are performed in the adiabatic limit. (a) The prepared state, $\Psi_\mathrm{p}$, is the (0,2) ground state and the separation  stage involves the avoided crossing I which is type (A); at the beginning of the evolution stage the system is in a (1,1) state. Due to the hyperfine dephasing during the evolution stage (e, $\tau_s$), states blocked for crossing I are populated. Those blocked states can still \emph{return to} $(0,2)$ at the joining stage due to the disorder-induced avoiding crossing II. The return probability thus depends on the detuning value at the measuring stage, see that $P^\mathrm{i}\smneq P^\mathrm{m}$, see Sec.\ref{SC:ResPrepGS}. (b) The prepared state is the (0,2) highest excited state (case shown is for $\Delta^{\mathrm{L}}<\Delta^{\mathrm{R}}$). At the separation stage the system passes through multiple Landau-Zener process, moreover, the first one is the disorder-induced avoided crossing, III'. The system is not guaranteed to be in a (1,1) state  at the beginning of the evolution stage, we study in detail this case in Sec.\ref{SC:ResPrepHS}.}
   \label{FG:Pinf1}
   \vspace{-0.2cm}
\end{figure*}

It is worth to note that in the low $t$ picture the knowledge of $\beta$ allows one to quantify all gaps in the 22 by 22 system as a function of a single one. We choose as reference the valley conserving avoided crossing I. In a clean system the energy gap, $2\sqrt{2} \left|t\right|$, is common to the six existing avoided crossings, namely, I, I' and the four crossings associated with case V.\cite{ReynosoFlensberg2011} Here one has twenty avoided crossings---I, I', III, III', II ($4$), II' ($4$), IV ($4$) and IV' ($4$)---and four distinct energy gaps,
\bese
\bea
\Delta_\mathrm{I} = \Delta_\mathrm{I'} &=&2\sqrt{2} \left|t_\mathrm{L -,R -}\right|,\\
\Delta_\mathrm{III} =  \Delta_\mathrm{III'} &=&\sqrt{\beta} \Delta_\mathrm{I}, \\
\Delta_\mathrm{IV} =  \Delta_\mathrm{IV'} &=&\sqrt{\frac{1}{2}} \Delta_\mathrm{I}, \\
\Delta_\mathrm{II} =  \Delta_\mathrm{II'} &=&\sqrt{\frac{\beta}{2}} \Delta_\mathrm{I}. \label{EQ:gapsBETAii}
\eea
\label{EQ:gapsBETA}
\eese
In the next section we make use of these relations for calculating the outcomes of the multiple Landau-Zener processes in the separation and joining stages.

\section{Return probability, method and results}
\subsection{Obtaining the return probability}

Having identified the complexity added by the presence of valley mixing in the double dot we now develop a scheme for computing the outcome of the return probability experiment, including all stages. The path we take is designed for gaining knowledge about the physics of the experiment with a degree of generality, any quantitative study designed for fitting a particular experiment must include further details as the actual pulse shapes and the dependence of the parameters with detuning. A full numeric time-dependent approach would be necessary if the avoided crossings overlap each other avoiding the application of the Landau-Zener formula separately for each crossing.

The most important information that the experiment provides is the dephasing time, $T_2^*$. It is also important the shape of $P(\tau_s)$ and in particular its saturation value,
\begin{equation}
P_\infty \equiv \lim_{\tau_s \rightarrow \infty} P(\tau_s).
\end{equation}
Our theoretical study aims for a qualitative understanding of which saturation values can be expected due to the hyperfine field induced dephasing in a disordered nanotube double dot. We show that the avoided crossing induced by disorder can also affect the return probability value even in the absence of dephasing. If the probability,
\begin{equation}
P_0 \equiv \lim_{\tau_s \rightarrow 0} P(\tau_s),
\end{equation}
is smaller than $1$, it means that the application of the separation and joining stages---without waiting at the evolution stage---is reducing the return probability due to the additional LZ processes. As it is discussed below, this statement is relevant for prepared states different than the ground state. Figure \ref{FG:Pinf1} sketches two situations studied below, preparing the ground state, panel (a), and preparing an excited state, panel (b).

We focus now on the preparation stage. Any given (0,2) prepared state can be decomposed into the six $(0,2)$ eigenstates of Eq.\eqref{EQ:02sols0En}, that we label $|l,(0,2)\rangle$, as, $|\Psi_\mathrm{p}\rangle=\sum_{l\smeq1}^6 a_l
|l,(0,2)\rangle$. To obtain the return probability one computes the average, over an ensemble of random
hyperfine fields, of the single-shot measurement's outcomes; the final result does not depend on the phases of coefficients $a_l$,\cite{ReynosoFlensberg2011}
\begin{equation}
P_{\psi_p}(\tau_s)= \sum_l|a_l|^2 P_{l}(\tau_s),
\label{EQ:Pprep}
\end{equation}
where $P_l(\tau_s)$ is the return probability found if the (0,2) eigenstate $|l,(0,2)\rangle$ is prepared. We then study separately the RPE assuming that a given eigenstate is prepared. Equation \eqref{EQ:Pprep} can also be applied to prepared mixed states, e.g. a thermal mixture, by replacing $|a_l|^2$ with the probability weight of the state $|l,(0,2)\rangle$ in the prepared mixture. However here we will assume that the $k_BT<\Delta^\mathrm{R/L}$ (for typical systems $\Delta^\mathrm{R}$ and $\Delta^\mathrm{L}$ are a few hundreds of $\mu\mathrm{eV}$ and the temperature is below $100\mathrm{mK}$, ($\kt\leq9\mu\mathrm{eV}$)) and therefore individual eigenstates can be prepared.\cite{ChurchillFlensberg2009}

When preparing (0,2) excited states, see Fig.\ref{FG:Pinf1}(b), the system must pass through multiple Landau-Zener processes at the separation stage and similarly at the joining stage. As shown in Fig.\ref{FG:Pinf1}(a), this is not the case if the ground state is prepared. In order to solve this problem for a general prepared state, we profit from the low $t$ picture and from the fact that the experiment is performed over an ensemble of hyperfine realizations. This allows us to obtain the final result by analyzing the Landau-Zener processes from a probabilistic perspective (free of interference effects) and combining those (both separation and joining stages) with the averaged behavior of the dephasing due to hyperfine at the evolution stage. It is important to keep in mind that $t$ is still much bigger than the hyperfine characteristic energy $\sigma_{\mathrm{H}}$ of Eq.\eqref{EQ:variances}: the duration of the separation and joining stages (see Fig.\ref{FG::f1}(c)) is shorter than $\hbar/\sigma_{\mathrm{ H}}$ and therefore it is still a good approximation to consider the effect of the hyperfine interaction only at the evolution stage.

When applying the probabilistic approach, we are assuming that the quantum interference effects, that arise only in Landau-Zener loops (when the system can arrive to a given state through different paths), average out. In order to explain the approximation we focus on the example shown in Fig.\ref{FG:Pinf1}(b) in which the highest excited (0,2) state is prepared. At the separation stage, the prepared state first visits the avoided crossing III' and then the crossing II' (note that I' and IV' are doublet conserving avoided crossings that in the figure are assumed as adiabatic processes). The energy difference between the two branches of the loop is given by $\Delta^\mathrm{L}$. If, during the separation detuning pulse, the time spent inside the loop, $T_l$, varies $\Delta T_l$, the phase difference that controls the interference after the crossing II' changes in $\Delta\phi \smap \Delta^\mathrm{L} \Delta T_l /\hbar$. The interference averages out if the phase gets randomized in a range $\Delta\phi> 2\pi$.

Taking the typical experimental values for $\Delta^\mathrm{L}$, the randomization happens if $\Delta T_l$ varies in a range of at least 15ps. Such $\Delta T_l$ variations are expected to be present in the experiment when combining the errors in preparing the system at point ``p" (see Fig.\ref{FG::f1}a) with the imperfections of the detuning rising pulse from ``p" to ``e". If, on the other hand, $\Delta T_l< 10$ps, the phase changes are not big enough and the interference effects would persist.\cite{NoriLZS2010} In such cases the single-shot cycles can be designed ad hoc with different times $T_l$ in order to get rid of the interference in the Landau-Zener loop and the treatment we do here would still be valid.\cite{DavidReillyPC2011} The main conclusions of the paper regarding the preparation of excited states are that the experiment no longer responds to the standard return probability experiment, leading to $P_0<1$. Those conclusions apply to the general case: we have performed full time dependent simulations (that contain the interference effects) and, if they are not tuned to sit in a destructive interference condition, they share this feature.

By working in the limit that the quantum interference in the Landau-Zener loops can be neglected, we manage to analyze in a simpler frame the general features that arise when preparing excited states. We define the matrices $M^{(\mathrm{s})}$ for the separation stage, $M^{(\mathrm{e})}(\tau_s)$ for the evolution stage, and $M^{(\mathrm{j})}$ for the joining stages. We make these matrices explicit below. They operate on the space of the 22 eigenstates with zero-tunneling, i.e., the six (0,2) states (that can be prepared) and the sixteen (1,1) states. The result of operating with these matrices on a given vector gives \emph{the probabilities} of finding the system in a new state after performing the stage. Therefore the vectors and matrices here are real positive numbers; in what follows, for each regular quantum state we associate a real probability vector
using the modified ket and bra notation,
\be
\ketLR{\psi} \rightarrow \ketLRr{\psi}.
\ee
We have introduced the~~$\widetilde{}$~~symbol to clearly distinguish the probabilistic picture from usual quantum mechanics. This non-quantum mechanical bra and ket notation greatly simplifies the writing of the matrix elements of the probability matrices and their operation on probability vectors.

Following Eq.\eqref{EQ:Pprep} $|a_l|^2$ are the nonzero components of $\ketLRr{\Psi_\mathrm{p}}$, the probability vector associated with the prepared state; the components lie in the (0,2) block of the vector. Since in what follows we study the RPE after preparation of specific (0,2) eigenstates then a single component of $\ketLRr{\Psi_\mathrm{p}}$ is nonzero and equal to 1. We apply the separation stage and obtain the real vector,
\be
\ketLRr{\Psi_\mathrm{e}^0}=M^{(\mathrm{s})} \ketLRr{\Psi_\mathrm{p}},
\ee
that contains the probabilities of finding the system in each of the 22 states at the beginning of the evolution stage. The matrix $M^{(\mathrm{e})}(\tau_s)$ is applied next, it contains the transitions probabilities between the different states due to the action of the hyperfine field during a time $\tau_s$. The average over the hyperfine field ensemble is already contained in $M^{(\mathrm{e})}(\tau_s)$. The probabilities after the evolution stage are,
\be
\ketLRr{\Psi_\mathrm{e}(\tau_s)}=M^{(\mathrm{e})}(\tau_s) \ketLRr{\Psi_\mathrm{e}^0}.
\label{EQ:Mevol}
\ee
And finally we apply the joining stage matrix to obtain the final probability vector,
\be
\ketLRr{\Psi_\mathrm{m}}=M^{(\mathrm{j})} \ketLRr{\Psi_\mathrm{e}(\tau_s)}.
\ee
The probability to measure the system in an (0,2) charge state is obtained simply by summing over the six components associated with (0,2) states. We define the (0,2) real vector proyector, $\ketLRr{(0,2)}$, having the six $(0,2)$ components equal to $1$ and zero for all the other ones. The return probability then becomes,
\be
P(\tau_s)\smeq   \braLNr{(0,2)} \ketLRr{\Psi_\mathrm{m}}.
\label{EQ:02Pprojection}
\ee

\subsubsection{Separation and joining stages}
\label{SC:MjMs}
 The states involved in each avoided crossing follow from the analysis presented in Sec.\ref{SC:classification}. From Table \ref{TB:charC} and Fig.\ref{FG:detun} we recognize four different detuning values, $\ve_\mathrm{I}$, $\ve_\mathrm{II}$, $\ve_\mathrm{III}$ and $\ve_\mathrm{IV}$ in which two avoided crossings operate. As an example, at $\ve_\mathrm{I}$ the actions of the avoided crossings $\mathrm{I}$ and $\mathrm{IV}$ are comprised in the matrix, $M^{(\mathrm{s})}_{\ve_\mathrm{I}}$, where $(\mathrm{s})$ refers to the separation stage. Since the crossing $\mathrm{I}$ is a Landau-Zener process between states $\ketLR{S_{--}^{(0,2)}}$ and $\ketLR{S_{--}^{(1,1)}}$ we have,
\bese
\bea
\braLRr{S_{--}^{(0,2)}}M^{(\mathrm{s})}_{\ve_\mathrm{I}}\ketLRr{S_{--}^{(0,2)}}&=&\braLRr{S_{--}^{(1,1)}}M^{(\mathrm{s})}_{\ve_\mathrm{I}}\ketLRr{S_{--}^{(1,1)}}=\overline{P}_\mathrm{I}^{(\mathrm{s})}, \\
\braLRr{S_{--}^{(0,2)}}M^{(\mathrm{s})}_{\ve_\mathrm{I}}\ketLRr{S_{--}^{(1,1)}}&=&\braLRr{S_{--}^{(1,1)}}M^{(\mathrm{s})}_{\ve_\mathrm{I}}\ketLRr{S_{--}^{(0,2)}}=P_\mathrm{I}^{(\mathrm{s})} . \eea
\eese
Similarly, given the mixing at crossing $\mathrm{IV}$ described by Eq.\eqref{EQ:HT2pmix}, we have,
\bese
\bea
\braLRr{S_{-+}^{(0,2)}}M^{(\mathrm{s})}_{\ve_\mathrm{I}}\ketLRr{S_{-+}^{(0,2)}}&=&\braLRr{S_{-+}^{(1,1)}}M^{(\mathrm{s})}_{\ve_\mathrm{I}}\ketLRr{S_{-+}^{(1,1)}}=\overline{P}_\mathrm{IV}^{(\mathrm{s})}, ~~~~~~~~~~~~~\\
\braLRr{T_{\sigma_t,-+}^{(0,2)}}M^{(\mathrm{s})}_{\ve_\mathrm{I}}\ketLRr{T_{\sigma_t,-+}^{(0,2)}}&=&\braLRr{T_{\sigma_t,-+}^{(1,1)}}M^{(\mathrm{s})}_{\ve_\mathrm{I}}\ketLRr{T_{\sigma_t,-+}^{(1,1)}}=\overline{P}_\mathrm{IV}^{(\mathrm{s})}, \\ \braLRr{S_{-+}^{(1,1)}}M^{(\mathrm{s})}_{\ve_\mathrm{I}}\ketLRr{S_{-+}^{(0,2)}}&=&\braLRr{S_{-+}^{(0,2)}}M^{(\mathrm{s})}_{\ve_\mathrm{I}}\ketLRr{S_{-+}^{(1,1)}}=P_\mathrm{IV}^{(\mathrm{s})},\\ \braLRr{T_{\sigma_t,-+}^{(1,1)}}M^{(\mathrm{s})}_{\ve_\mathrm{I}}\ketLRr{T_{\sigma_t,-+}^{(0,2)}}&=&\braLRr{T_{\sigma_t,-+}^{(0,2)}}M^{(\mathrm{s})}_{\ve_\mathrm{I}}\ketLRr{T_{\sigma_t,-+}^{(1,1)}}=P_\mathrm{IV}^{(\mathrm{s})},~~~~~~
\eea
\eese
with $\sigma_t\smeq\{-1,0,1\}$. All the remaining matrix elements of $M^{(\mathrm{s})}_{\ve_\mathrm{I}}$ are trivial, being 0 the non-diagonal ones and $1$ the diagonals ones. We construct in the same way the remaining three matrices of the separation stage, $M^{(\mathrm{s})}_{\ve_\mathrm{II}}$, $M^{(\mathrm{s})}_{\ve_\mathrm{III}}$, and $M^{(\mathrm{s})}_{\ve_\mathrm{IV}}$. The matrices for the joining stage follow from replacing the LZ-probabilities with the ones corresponding to the joining stage $P^{(\mathrm{j})}$. These probabilities can be different because, even though the gaps of the avoiding crossings are the same, we allow for a difference in the detuning rate of change between the separation and the joining stages. We define the ratio between the speeds as,
\be
\kappa\smeq \frac {v^{(\mathrm{j})}}{v^{(\mathrm{s})}}.
\label{EQ:kappa}
\ee
For simplicity we assume that the detuning speeds $v^{(\mathrm{s})}$ and $v^{(\mathrm{j})}$ do not depend on detuning fast enough to be considered different at the four relevant detunings (i.e., in the detuning range $[\ve_\mathrm{III},\ve_\mathrm{II}]$, which from Table \ref{TB:charC} is $[-(\Delta^\mathrm{L}+\Delta^\mathrm{R})/2,(\Delta^\mathrm{L}+\Delta^\mathrm{R})/2]$). One can write all the Landau-Zener probabilities as a function of a single probability, for instance $P_\mathrm{I}^{(\mathrm{s})}$, once $\kappa$ and $\beta$ are known. We do so by combining the Landau-Zener formula of Eq.\eqref{EQ:LZformula}, the relations fixed by $\beta$ between all the gaps in Eq.\eqref{EQ:gapsBETA} and the definition of $\kappa$ given in Eq.\eqref{EQ:kappa}.

The full separation (joining) stage matrix is then given by the successive application of the matrices above in decreasing (increasing) detuning order,
\bese
\bea
M^{(\mathrm{s})}= \begin{cases}
M^{(\mathrm{s})}_{\ve_\mathrm{III}}M^{(\mathrm{s})}_{\ve_\mathrm{IV}}M^{(\mathrm{s})}_{\ve_\mathrm{I}}M^{(\mathrm{s})}_{\ve_\mathrm{II}} & \text{for $\Delta^\mathrm{L}>\Delta^\mathrm{R}$} \\
M^{(\mathrm{s})}_{\ve_\mathrm{III}}M^{(\mathrm{s})}_{\ve_\mathrm{I}}M^{(\mathrm{s})}_{\ve_\mathrm{IV}}M^{(\mathrm{s})}_{\ve_\mathrm{II}} & \text{for $\Delta^\mathrm{L}<\Delta^\mathrm{R}$}
\end{cases},\\
M^{(\mathrm{j})}= \begin{cases}
M^{(\mathrm{j})}_{\ve_\mathrm{II}}M^{(\mathrm{j})}_{\ve_\mathrm{I}}M^{(\mathrm{j})}_{\ve_\mathrm{IV}}M^{(\mathrm{j})}_{\ve_\mathrm{III}} & \text{for $\Delta^\mathrm{L}>\Delta^\mathrm{R}$} \\
M^{(\mathrm{j})}_{\ve_\mathrm{II}}M^{(\mathrm{j})}_{\ve_\mathrm{IV}}M^{(\mathrm{j})}_{\ve_\mathrm{I}}M^{(\mathrm{j})}_{\ve_\mathrm{III}} & \text{for $\Delta^\mathrm{L}<\Delta^\mathrm{R}$}
\end{cases}.
\eea
\label{EQ:MjMs}
\eese

\subsubsection{Evolution stage, hyperfine interaction}
\label{SC:Me}
As discussed in Sec.\ref{SC:hyperfine}, during each single-shot cycle the hyperfine fields can be considered fixed. Because $\Delta^{\xi}\gg \sigma_\mathrm{H}$, $\xi=L,R$, the evolution of an electron is assumed to be restricted to the Kramers doublet it occupies, i.e., the dynamics follows Hamiltonians $H^\mathrm{L d_L}_{\tiny\rm{HF}}$ or $H^\mathrm{R d_R}_{\tiny\rm{HF}}$ as the one given in Eq.\eqref{EQ:HhfiKds}. The phase prefactors incorporated in the unitary transformation in Eq.\eqref{EQ:newEig} modify the effective fields presented in Eq.\eqref{EQ:effHfine}. This does not introduce any complication because the transformed Hamiltonians have the form of Eq.\eqref{EQ:HhfiKds} with transformed hyperfine field components that follow the same zero mean Gaussian distributions of Eq.\eqref{EQ:variances2}; i.e., the effective hyperfine field seen in any given Kramers doublet retains its statistical isotropic property.

Therefore, the time evolution operator (taken from $t_0\smeq0$) for an electron in the doublet $(\xi \mathrm{d}_\xi)$ is,
\bea
u^{\xi \mathrm{d}_\xi}(t)&=&{\rm exp}\left(-\ci H^{\xi \mathrm{d}_\xi}_{\tiny\rm{HF}} t \right) \nonumber \\
&=&\cos(\omega^{\xi \mathrm{d}_\xi} t)\sigma^{\xi \mathrm{d}_\xi}_0 - \ci \sin(\omega^{\xi \mathrm{d}_xi} t)~ \boldsymbol{ \sigma}^{\xi \mathrm{d}_\xi}\cdot \hat{{\bf n}}^{\xi \mathrm{d}_\xi},
\label{EQ:evol1part}
\eea
where $\boldsymbol{ \sigma}^{\xi\mathrm{d}_\xi}\smequiv(\sigma^{\xi\mathrm{d}_\xi}_x,\sigma^{\xi\mathrm{d}_\xi}_y,\sigma^{\xi\mathrm{d}\xi}_z)$ are the Pauli matrices operating on the pseudo-spin of the $(\xi \mathrm{d}_\xi)$ Kramers doublet, $B^{\xi \mathrm{d}_\xi} \hat{{\bf n}}^{\xi\mathrm{d}_\xi} \smequiv {\bf B}^{\xi \mathrm{d}_\xi}$ is the effective hyperfine field vector with $\hat{{\bf n}}^{\xi\mathrm{d}_\xi}$ being the unit vector that defines the direction of the field, and $\omega^{\xi\mathrm{d}_\xi}\smeq B^{\xi\mathrm{d}_\xi}/\hbar$, the precession frequency.

In order to compute the dynamics for any (1,1) or (0,2) state we use the Slater determinants' basis constructed with eigenstates of the two quantum dots, $\ketLR{\xi\mathrm{d}_\xi\sigma_\xi}$. The time evolution of each Slater determinant is dictated by,
\bea
U(t)  \left|{}_{\xi \mathrm{d}_\xi \sigma_\xi}^{\xi' \mathrm{d}_{\xi'} \sigma_{\xi'}}\right> &=& \sum_{\sigma_1,\sigma_2} u^{\xi\mathrm{d}_{\xi}}_{\sigma_1,\sigma_{\xi}}(t) \times u^{\xi'\mathrm{d}_{\xi'}}_{\sigma_2,\sigma_{\xi'}}(t)
\left|{}_{\xi \mathrm{d}_\xi \sigma_1}^{\xi' \mathrm{d}_{\xi'} \sigma_{2}}\right>,
\nonumber \\
&=& \sum_{\sigma_1,\sigma_2} U^{\xi\mathrm{d}_{\xi};\xi'\mathrm{d}_{\xi'}}_{\sigma_1,\sigma_{\xi};\sigma_2,\sigma_{\xi'}}(t) \left|{}_{\xi \mathrm{d}_\xi \sigma_1}^{\xi' \mathrm{d}_{\xi'} \sigma_{2}}\right>.
\label{EQ:TimeEvolOp0}
\eea
From Eq.\eqref{EQ:Mevol} the matrix element $P_{i,f}(\tau_s)\smeq \braLRr{\Phi_f} M^{(\mathrm{e})}(\tau_s)\ketLRr{\Phi_i}$ is the probability (averaged over all the single-shot cycles) of finding the system after a time $\tau_s$ in the state $\ketLR{\Phi_f}$ given that the state at the beginning of the evolution stage is $\ketLR{\Phi_i}$. From the doublet conservation dynamics dictated by Eq.\eqref{EQ:evol1part}, it follows that nonzero matrix elements arise only if the occupied Kramers doublets---for example $\xi_1 \mathrm{d}_1$ and $\xi_2 \mathrm{d}_2$---in $\ketLR{\Phi_i}$ and in $\ketLR{\Phi_f}$ are the same.

The probabilities $P_{i,f}(\tau_s)$ are obtained by applying the time evolution operator of Eq.\eqref{EQ:TimeEvolOp0} to the two-particle state $\ketLR{\Phi_i}$, and averaging $\left|\braLR{\Phi_f}U(\tau_s)\ketLR{\Phi_i}\right|^2$ over the hyperfine field realizations that are independent in different dots and doublets. As shown in Sec.\ref{SC:hyperfine}, for zero magnetic field the effective hyperfine field components in a given dot follow zero mean Gaussian distributions with the same variances irrespective of the Kramers doublet, i.e., $\sigma_{\xi} \smequiv \sigma_{\xi \mathrm{d}_\xi,j}$ with $j\smeq x,y,z$ and $\mathrm{d}_\xi\smeq\pm$. All the probabilities can be expressed in terms of only two ensemble averages per dot,\cite{MerkulovHyperfine,CoishDQD}
\bea
f^{\xi}_C(\tau_s)&\equiv&\braket{\cos^2(\omega^{\xi\mathrm{d}_{\xi}} \tau_s)}_{\rm hyperfine}\nonumber \\
 &=& \frac{1}{2}\left[1 +  \left(1 - 4 \frac{\sigma_{\xi}^2 \tau_s^2}{\hbar^2}  \right){\exp}{\left(-2 \frac{\sigma_{\xi}^2 \tau_s^2}{\hbar^2} \right)}\right], \\
f^{\xi}_S(\tau_s)&\equiv&\braket{\left(n^{\xi\mathrm{d}_{\xi}}_j\right)^2\sin^2(\omega^{\xi\mathrm{d}_\xi} \tau_s)}_{\rm hyperfine}\nonumber \\
 &=& \frac{1}{6}\left[1 -  \left(1 - 4 \frac{\sigma_{\xi}^2 \tau_s^2}{\hbar^2}  \right)\exp{\left(-2 \frac{\sigma_{\xi}^2 \tau_s^2}{\hbar^2} \right)}\right].
\eea
Note that the average over the hyperfine fields of $\left(n^{\xi\mathrm{d}_\xi}_j\right)^2\sin^2(\omega^{\xi\mathrm{d}_xi}\tau_s)$---where $n^{\xi\mathrm{d}_\xi}_j$ is a Cartesian component of the unit vector $\hat{{\bf n}}^{\xi\mathrm{d}_\xi}$---is independent of $j$ because of the isotropy, in average, of the effective hyperfine field. Due to the zero mean of the Gaussian distributions that govern all components of ${\bf B}^{\xi \mathrm{d}_\xi}$, terms having odd powers in any given component $n^{\xi\mathrm{d}_\xi}_j$ average out being absent in $P_{i,f}(\tau_s)$.

The nonzero probability elements for the evolution depend on $\tau_s$ as sums of products of the above functions. Starting with spin polarized states the matrix elements are,
\bese
\bea
\braLRr{T^{(N_\mathrm{L},N_\mathrm{R})}_{\pm 1,\mathrm{d_1 d_2}}} M^{(\mathrm{e})}\ketLRr{T^{(N_\mathrm{L},N_\mathrm{R})}_{\pm 1,\mathrm{d_1 d_2}}}&=&(f_C^{\xi_1} +f_S^{\xi_1})(f_C^{\xi_2} +f_S^{\xi_2}),~~~~~~~\\
\braLRr{T^{(N_\mathrm{L},N_\mathrm{R})}_{\mp 1,\mathrm{d_1 d_2}}} M^{(\mathrm{e})}\ketLRr{T^{(N_\mathrm{L},N_\mathrm{R})}_{\pm 1,\mathrm{d_1 d_2}}}&=&4 f_S^{\xi_1} f_S^{\xi_2},~~~~~~~\\
\braLRr{T^{(N_\mathrm{L},N_\mathrm{R})}_{0,\mathrm{d_1 d_2}}} M^{(\mathrm{e})}\ketLRr{T^{(N_\mathrm{L},N_\mathrm{R})}_{\pm 1,\mathrm{d_1 d_2}}}&=&2 f_S^{\xi_1} f_S^{\xi_2} + f_S^{\xi_1} f_C^{\xi_2}+f_C^{\xi_1} f_S^{\xi_2},~~~~~~~\\
\braLRr{S^{(N_\mathrm{L},N_\mathrm{R})}_{\mathrm{d_1 d_2}}} M^{(\mathrm{e})}\ketLRr{T^{(N_\mathrm{L},N_\mathrm{R})}_{\pm 1,\mathrm{d_1 d_2}}}&=&2 f_S^{\xi_1} f_S^{\xi_2} + f_S^{\xi_1} f_C^{\xi_2}+f_C^{\xi_1} f_S^{\xi_2},~~~~~~~~
\eea
\eese
while if the evolution starts in singlet like states we have,
\bese
\bea
\braLRr{S^{(N_\mathrm{L},N_\mathrm{R})}_{\mathrm{d_1 d_2}}} M^{(\mathrm{e})}\ketLRr{S^{(N_\mathrm{L},N_\mathrm{R})}_{\mathrm{d_1 d_2}}}&=& f_C^{\xi_1}f_C^{\xi_2} +3 f_S^{\xi_1} f_S^{\xi_2},~~~~~~~
\label{EQ:singletEVOL}\\
\braLRr{T^{(N_\mathrm{L},N_\mathrm{R})}_{0,\mathrm{d_1 d_2}}} M^{(\mathrm{e})}\ketLRr{S^{(N_\mathrm{L},N_\mathrm{R})}_{\mathrm{d_1 d_2}}}&=& f_C^{\xi_1}f_S^{\xi_2}+f_S^{\xi_1}f_C^{\xi_2} +2 f_S^{\xi_1} f_S^{\xi_2},~~~~~~~~~\\
\braLRr{T^{(N_\mathrm{L},N_\mathrm{R})}_{\pm,\mathrm{d_1 d_2}}} M^{(\mathrm{e})}\ketLRr{S^{(N_\mathrm{L},N_\mathrm{R})}_{\mathrm{d_1 d_2}}}&=& f_C^{\xi_1}f_S^{\xi_2}+f_S^{\xi_1}f_C^{\xi_2} +2 f_S^{\xi_1} f_S^{\xi_2},~~~~~~~~
\eea
\label{EQ:MevolSinglet}
\eese
where for $(N_\mathrm{L},N_\mathrm{R})\smeq(1,1)$ there are four possible $\mathrm{d_1 d_2}$ cases with $\xi_1\smeq\mathrm{L}$ and $\xi_2\smeq\mathrm{R}$, while for $(N_\mathrm{L},N_\mathrm{R})\smeq(0,2)$, $\xi_{1,2}\smeq\mathrm{R}$, and the only valid case is $\mathrm{d_1 d_2}\smeq+-$.
The matrix elements of $M^{(\mathrm{e})}(\tau_s)$ for the case that the evolution starts in a $T_0$-like function can be obtained from Eq.\eqref{EQ:MevolSinglet} by making the replacement $T_0 \leftrightarrow  S$. Two nonzero matrix elements remain to be defined, those for (0,2) singlet-like states in the same Kramers doublet. Since these are nondegenerated states that are well separated in energy from the remaining 21 states, we have,
\be
\braLRr{S^{(0,2)}_{\mathrm{d d}}} M^{(\mathrm{e})}\ketLRr{S^{(0,2)}_{\mathrm{d d}}}=1.
\label{EQ:02singletsTrivial}
\ee

\subsection{Results, zero magnetic field}
\label{SC:Results}
The separation and joining stages produce different outcomes for singlet-like and triplet-like states even if they belong to the same subset $(\xi_1\mathrm{d}_1,\xi_2\mathrm{d}_2)$. This can be seen in Fig.\ref{FG:cross}(a) for crossings I, I', III and III' where only the singlet-likes states are mixed. It is therefore important to see how the probabilities of finding the system in the different states behave for long times after the evolution stage.

First, from the results in Sec.\ref{SC:Me}, we see that $P_{i,i}(0)\smeq 1$, with $P_{i,i}(\tau_s)$ the  probability of finding the system after the evolution in the same state as the initial state, $\ketLRr{\Phi_i}$. In general, $P_{i,i}(\tau_s)$ decays to a saturation value in a time of the order $\hbar/\sigma_\mathrm{H}$; this decaying time is sensitive to differences in the number of ${}^{13}$C atoms in each quantum dot leading to $\sigma^\mathrm{L}_\mathrm{H}\smneq\sigma^\mathrm{R}_\mathrm{H}$. On the other hand the saturation value is robust because the limits $f_C^{\xi}(\tau_s\rightarrow\infty)\smeq 1/2$ and $f_S^{\xi}(\tau_s\rightarrow\infty)\smeq 1/6$ are independent of the variances as long as they are finite. For example, when $\ketLR{\Phi_i}$ is singlet-like (excluding the trivial $P_{i,i}(\tau_s)\smeq 1$ cases of Eq.\eqref{EQ:02singletsTrivial}) we get from Eq.\eqref{EQ:singletEVOL} that $P_{i,i}(\infty)=1/3$. This means that the system is found with probability 2/3 in triplet-like states that will produce different outcomes after the joining stage.

In summary, the evolution matrix is robust in two limits, at $\tau_s\rightarrow\infty$ we get the saturation behavior while $M^{(\mathrm{e})}(\tau_s\smeq 0)$ is just the identity matrix. For this reason we focus on $P_0$ and $P_\infty$. In the following we discuss in detail the physics of the most interesting situations, those shown in Fig.\ref{FG:Pinf1}. As discussed above, with the knowledge of $\kappa$, $\beta$ and any one of the LZ probabilities we can fully describe the separation and joining stages $M^{(\mathrm{s})}$ and $M^{(\mathrm{j})}$. While the disorder imposes the value of $\beta$, the chosen detuning rates, $v^{(\mathrm{j})}$ and $v^{(\mathrm{s})}$, control both our reference probability (usually a LZ probability in a specific avoided crossing that we desire to maintain in the adiabatic limit) and the value of $\kappa$.

\begin{figure*}[!ht]
   \centering
   \includegraphics[width=.85\textwidth]{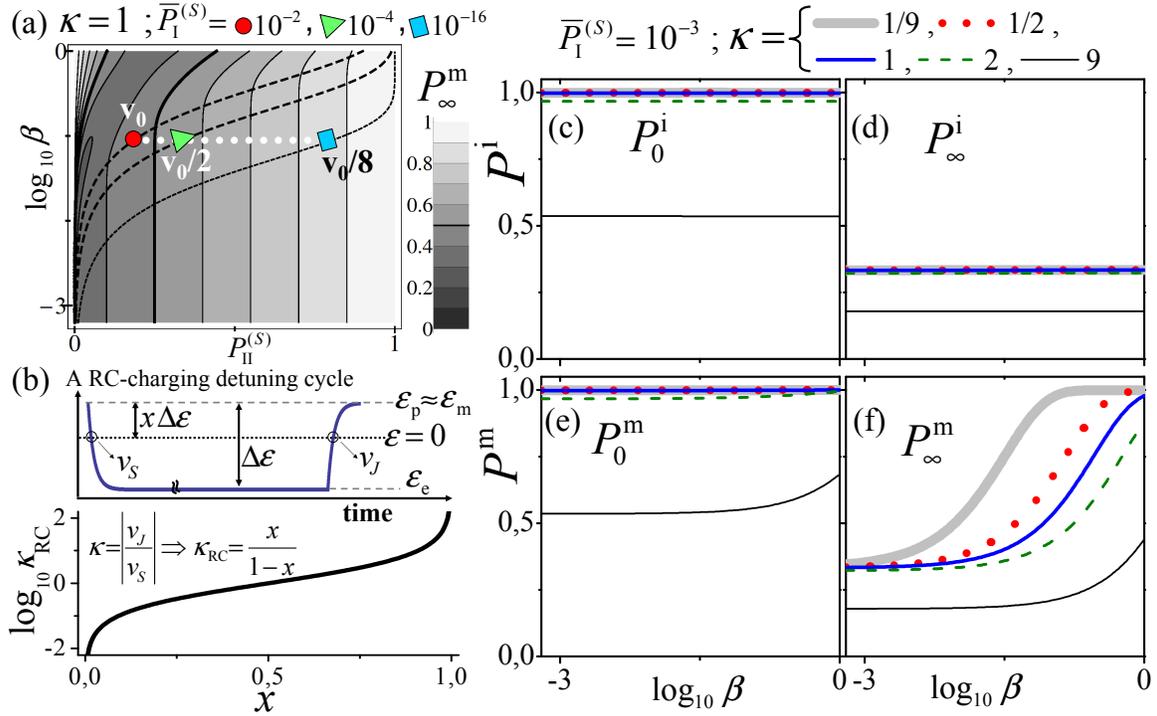}
   \vspace{-0.3cm}
   \caption{Return probability if the (0,2) ground state, $\ketLR{S_{--}^{(0,2)}}$, is prepared [see Fig\ref{FG:Pinf1}(a)]. (a) Saturation return probability $P_\infty^\mathrm{m}$ at the measurement region ($\ve_\mathrm{m}>\ve_\mathrm{II}$) as a function \emph{of} $\beta$ and \emph{of} the LZ probability in the DFAC II, $P^{(\mathrm{s})}_\mathrm{II}$, for $\kappa\smeq 1$. The three trajectories fall in the experimental relevant regime: the LZ probability at I is assured to be in the adiabatic limit, i.e., $\overline{P}^{(\mathrm{s})}_\mathrm{I}$ is small. The detuning velocity $v_0$ is chosen such that $\overline{P}^{(\mathrm{s})}_\mathrm{I}\smeq 0.01$. As the speed decreases  ($v_0/2$ and $v_0/8$) II becomes more effective; both $P^{(\mathrm{s})}_\mathrm{II}$ and $P_\infty^\mathrm{m}$ grow. (b) Example of $\kappa\smneq 1$ for detuning pulses generated using a RC-charging/discharging circuit. If the zero detuning region is closer to the preparation point than to the evolution point (this is $x<1/2$), a $\kappa_\mathrm{RC}<1$ situation is expected. Return probabilities along a physical relevant trajectory ($P^{(\mathrm{s})}_\mathrm{leak} \smeq \overline{P}^{(\mathrm{s})}_\mathrm{I}\smeq 10^{-3}$) are show as a function of $\beta$ and $\kappa$ in panels (c) and (d) [at the intermediate region, $\ve_\mathrm{I}>\ve_\mathrm{m}>\ve_\mathrm{II}$], and (e) and (f) [at the measurement region]. The values for $\tau_s\smeq 0$ (panels (c) and (e)) are $\beta$-independent and they remain close to $1$ unless the joining stage becomes too rapid ($\kappa\smeq 9$, see text). Panels (d) and (f) show the \emph{saturation} return probabilities. In (d) the joining stage does not include the crossing II and $P^{\mathrm{i}}_\infty \smap 1/3$ independently of $\beta$ and $\kappa$ (for $\kappa\leq 2$). In (f) the system passes through the crossing II before entering the measuring region and $P^{\mathrm{m}}_\infty > 1/3$. The return probability in (f) grows as the triplet-like states can return to (0,2) more effectively, i.e., the bigger it is $\beta$ and the slower is the joining stage.}
   \label{FG:GS}
   \vspace{-0.2cm}
\end{figure*}

The most natural choice for $\kappa$ is 1, when assuming that the detuning speeds at the separation and joining stages are the same. However that might not be the case experimentally. As an example, in Fig.\ref{FG:GS}(b) we show that pulse generation through the charging and discharging in a resistor-capacitor (RC) circuit can induce a $\kappa\smneq 1$ situation. This happens for all cases if the boundary between the (0,2) region and the (1,1) region [the detuning zone containing the avoided crossings, $\ve\smap 0$, or more specifically $\ve_\mathrm{III}>\ve>\ve_\mathrm{II}$] is not centered between the preparation and measurement points (see Fig.\ref{FG::f1}(a)). Experimentally the point ``p" is often closer to the boundary:\cite{Petta2005,ChurchillFlensberg2009} as shown in the picture, the separation stage is faster than the joining stage leading to $\kappa<1$. This means that all the avoided crossings would become \emph{more effective} at the joining stage and could, therefore, modify the outcome of the experiment. As we see in the following, a situation with $\kappa\smneq 1$ here has more impact on the final result than in 2DEG-based double dots where there is a unique energy gap due to interdot tunneling.

\label{SC:ResPrepGS}

\subsubsection{Return probability when preparing the (0,2) ground state}
Here we study the return probability if the ground state is prepared, $\ketLR{\Psi_p}\smeq\ketLR{S_{--}^{(0,2)}}$. The goal is understanding how the return probability experiment is affected by the doublet-flipping interdot tunneling, i.e., for $\beta>0$. We also investigate the effects of a mismatch in the rising times of the separation and the joining stages: the behavior for different values of $\kappa$. We start by choosing equal separation and joining detuning speeds, $\kappa\smeq 1$. For any given value of $\beta$ the speed chosen, $v^{(\mathrm{s})}\smeq v^{(\mathrm{j})}$, changes the outcome of the experiment. From Fig.\ref{FG:Pinf1}(a), we see that the probability of state conversion at the avoided crossing II (that arises due to the disorder) is meaningful for the experiment, therefore, instead of showing the results as a function of $v^{(\mathrm{j})}$ and $\beta$ we do it as a function of $P^{(\mathrm{s})}_\mathrm{II}$ and $\beta$. The result presented in Fig.\ref{FG:GS}(a) is the saturation return probability at the measurement region, $P^{\mathrm{m}}_\infty$; i.e., after a joining stage that takes the system to a large measurement detuning, $\ve_\mathrm{m}$, so that $\ve_\mathrm{m}>\ve_\mathrm{II}$.

In order to study a region in the two-dimensional space $\left(P^{(\mathrm{s})}_{\mathrm{II}},\beta\right)$ meaningful for the experiment we need an additional constrain: we will assume that at the time of performing the experiment one assures that the doublet conserving avoided crossing I, having the biggest gap in the system (for $\beta<1$), is in the adiabatic limit. As seen in Fig.\ref{FG:Pinf1}(a) this guarantees that the separation stage is fully effective leaving the system in the (1,1) state, $\ketLR{S_{--}^{(1,1)}}$. The condition for adiabatic separation at I  is equivalent to requiring a small complementary probability or leakage, $P^ {(\mathrm{s})}_{\mathrm{leak}}\smequiv \overline{P}^{(\mathrm{s})}_{\mathrm{I}} \smap 0$. From the LZ formula of Eq.\eqref{EQ:LZformula} and from the relations between the different gaps given in Eq.\eqref{EQ:gapsBETAii}, we obtain
\be
P^{(\mathrm{s})}_{\mathrm{II}}\left(\beta,P^{(\mathrm{s})}_{\mathrm{leak}}\right)= 1-\left(P^{(\mathrm{s})}_{\mathrm{leak}}\right)^{\frac{\beta}{2}}.
\ee
In Fig.\ref{FG:GS}(a) we present three curves $P^{(\mathrm{s})}_{\mathrm{II}}\left(\beta,P^{(\mathrm{s})}_{\mathrm{leak}}\right)$ with I approaching the adiabatic regime as $P^{(\mathrm{s})}_{\mathrm{leak}}\smeq 10^{-2},10^{-4}$ and $10^{-16}$. It is clear that for reducing $P^{(\mathrm{s})}_{\mathrm{leak}}$ one needs to slow down the detuning speed as,
\be
\left.v^{(\mathrm{s})}\right|_\mathrm{new} =\frac{\left. v^{(\mathrm{s})}\right|_\mathrm{old}}{y}~~\Rightarrow~~\left.P^{(\mathrm{s})}_{\mathrm{leak}}\right|_\mathrm{new} =\left(\left. P^{(\mathrm{s})}_{\mathrm{leak}}\right|_\mathrm{old}\right)^{y}.
\label{EQ:detuningSpeed0}
\ee
With $v_0$ being the detuning speed that assures $P^{(\mathrm{s})}_{\mathrm{leak}}\smeq 10^{-2}$, the cases with $P^{(\mathrm{s})}_{\mathrm{leak}}\smeq 10^{-4}$ and $ 10^{-16}$ shown in the figure are performed at speeds $v_0/2$ and $v_0/8$, respectively. The smaller $P^{(\mathrm{s})}_{\mathrm{leak}}$, the more the experimental curve shifts to the left; such a shift increases the effectiveness of the avoided crossing II leading to a larger return probability.

Having identified the experimentally relevant trajectories, we compute the probability of finding the system in a (0,2) state, for $\tau_s\smeq 0$ and for $\tau_s\rightarrow \infty$, choosing $P^{(\mathrm{s})}_{\mathrm{leak}}\smeq 10^{-3}$. The results, in general, only depend on $\beta$ and $\kappa$. As described above, a regular joining stage moves the detuning all the way to the $\ve_\mathrm{m}>\ve_\mathrm{II}$ region, i.e., the measurement region; this leads us to the probabilities $P^{\mathrm{m}}_0$ and $P^{\mathrm{m}}_\infty$. However, as sketched in Fig.\ref{FG:Pinf1}(a), one can also envision an experiment with a joining stage having a final detuning in the \emph{intermediate} region given by $\ve_\mathrm{I}<\ve_\mathrm{m}<\ve_\mathrm{II}$ and measuring the probabilities $P^{\mathrm{i}}_0$ and $P^{\mathrm{i}}_\infty$. We obtain the intermediate final state simply by excluding $M^{(\mathrm{j})}_{\ve_\mathrm{II}}$ in the composition of $M^{(\mathrm{j})}$ given in Eq.\eqref{EQ:MjMs}; the return probabilities are found applying Eq.\eqref{EQ:02Pprojection}, i.e., summing the probabilities of ending in an (0,2) state.

Here this distinction is very useful, because by studying how the $P^{\mathrm{m}}$ probabilities differ from the $P^{\mathrm{i}}$ ones, we manage to isolate the effect of the avoided crossing II. Moreover, the return probabilities $P^{\mathrm{i}}_0$ and $P^{\mathrm{i}}_\infty$ [as shown in Figs.\ref{FG:GS}(c) and (d)] behave as the well known spin-only double dot which gives $P^{\mathrm{i}}_0\smeq1$ and $P^{\mathrm{i}}_\infty \smeq 1/3$.\cite{MerkulovHyperfine,CoishDQD,Petta2005,TaylorDephasingTheory2007} This result is expected because, as we have shown above, the hyperfine field in the space of each Kramers doublet (for each dot) is  isotropic in average, and also because the behavior of crossing I is completely analogous to the crossing involving the (1,1) and the (0,2) spin-singlets in 2DEGs' dots. The independence with $\beta$ of the $P^{\mathrm{i}}$ probabilities is trivial because, for a given value of $\beta$ at the separation stage, the detuning speed is adjusted to reduce the leakage making the avoided crossing I adiabatic, and because the reduced joining stage does not involve the crossing II.

Regarding the dependence with $\kappa$ of the probabilities $P^{\mathrm{i}}$, we find a drastic reduction when the joining stage is performed too fast (see for example $\kappa\smeq9$). This means that the avoided crossing I becomes non-adiabatic and the (1,1) singlet-like state is not effectively converted to a (0,2) singlet-like state. By a reasoning similar to the one leading to Eq.\eqref{EQ:detuningSpeed0} one gets,
\be
\text{Because~}\kappa =\frac{ v^{(\mathrm{j})}}{v^{(\mathrm{s})}}~~\Rightarrow~~P^{(\mathrm{j})}_{\mathrm{leak}}=\left( P^{(\mathrm{s})}_{\mathrm{leak}}\right)^{\frac{1}{\kappa}}.
\label{EQ:detuningSpeed1}
\ee
Having $P^{(\mathrm{s})}_{\mathrm{leak}}\smeq 10^{-3}$ we obtain that $P^{(\mathrm{j})}_{\mathrm{leak}}\smap 0.464$ for $\kappa\smeq 9$, since the leaking at I becomes very high such too rapid joining stages are always avoided when designing the experiment. In what follows we concentrate the discussion on the smaller values of $\kappa$.

We note that for $\kappa\smlt 2$ the $P^{\mathrm{i}}$ values are virtually unaffected by changing the detuning speed of the joining stage (either decreasing or increasing it). On the other hand, the $P^{\mathrm{m}}_\infty$ probability [see Fig.\ref{FG:GS}(f)] is always affected by $\kappa$ when the double-dot system have $\beta\smgt 0.001$. The smaller $\kappa$, the larger $P^{\mathrm{m}}_\infty$ becomes, increasing the difference with the standard case of $P^{\mathrm{i}}_\infty\smap 1/3$.   Those effects are due to the avoiding crossing II because it provides a new path \emph{for returning to} (0,2) to the (1,1) triplet-like states that are blocked at the avoided crossing I [states in Eq.\eqref{EQ:triplets11} taking $\mathrm{d}\smeq \mathrm{d'}\smeq-$]. Of course, the enhancement of the return probability grows with $\beta$ even for $\kappa\smeq 1$. Assuming that I is adiabatic also at the joining stage (which is a good approximation for $\kappa\smlt 2$) we can write the return probability simply as,
\bese
\bea
P^{\mathrm{i}}\left(\tau_s\right)&=&f_C^{\mathrm{L}}\left(\tau_s\right)f_C^{\mathrm{R}}\left(\tau_s\right) +3 f_S^{\mathrm{L}}\left(\tau_s\right) f_S^{\mathrm{R}}\left(\tau_s\right),\\
P^{\mathrm{m}}\left(\tau_s\right)&=& P^{\mathrm{i}}\left(\tau_s\right)  + \left(1-P^{\mathrm{i}}(\tau_s)\right) P^{(\mathrm{j})}_{\mathrm{II}}.
\eea
\eese
The return probability $P^{\mathrm{i}}\left(\tau_s\right)$ is just Eq.\eqref{EQ:singletEVOL}, the well known probability for spin-only double dots at zero magnetic field of being found in a singlet-like state having started the evolution in the same singlet-like state. The enhancement term in $P^{\mathrm{m}}\left(\tau_s\right)$ is proportional to $P^{(\mathrm{j})}_{\mathrm{II}}$ and to the probability of being found after the evolution stage in a (1,1) triplet-like state, [$1-P^{\mathrm{i}}\left(\tau_s\right)$]. Experimentally, it is possible to explore if the system responds to this description by designing a joining pulse with different $v^{(\mathrm{j})}$ at the two avoided crossings. This can be done for instance by fixing $v^{(\mathrm{j})}$ at I slow enough to maintain $P^{(\mathrm{j})}_{\mathrm{leak}}$ small and by repeating the experiment for different values of $v^{(\mathrm{j})}$ at the avoiding crossing II.

To summarize this part, when the prepared state is the ground state the return probability experiment produces a very similar output to the case of a spin-only double dot. A mapping between the two situations in the absence of valley was done in our previous publication.\cite{ReynosoFlensberg2011} Here the hyperfine dynamics remains isotropic in average but the joining stage is affected by a disorder induced avoided crossing (II) that allows for an enhancement of the return probability because it unblocks the triplet-like (1,1) states.

\subsubsection{Return probability when preparing (0,2) higher energy states}
\label{SC:ResPrepHS}

The purpose of this section is to demonstrate the complications that arise when the prepared state is not the (0,2) ground state. We show that the simple procedure followed in Sec.\ref{SC:ResPrepGS} of reducing the leakage probability in the avoided crossings with the biggest gaps (for $\beta\leq 1$ the doublet conserving, I and I') does not assure a successful separation stage. After a successful separation stage, the system is in a (1,1) configuration, therefore, the probability of finding the system in (0,2) at the evolution region, $P^{\mathrm{e}}$, must be close to zero.

The situation we study is sketched in Fig.\ref{FG:Pinf1}(b), the prepared state is the highest energy excited state, $\ketLR{\Psi_p}\smeq\ketLR{S_{++}^{(0,2)}}$. As opposite to the ground state case, the separation stage first involves the LZ process in III', which is an avoided crossing induced by disorder (a DFAC). From Eq.\eqref{EQ:gapsBETA}, we have that $\Delta_\mathrm{III'} \smeq \sqrt{\beta} \Delta_\mathrm{I'}$, and this links the probabilities of state conversion in I' and III' (assuming the detuning speed is the same) as shown in Fig.\ref{FG:Mfac1}(b). For weak disorder $\Delta_\mathrm{III'}$ is small and therefore the adiabatic condition cannot be met while still maintaining the passage time larger than the characteristic time of the HFI. Instead, we explore the effect of III' being a non-adiabatic process in general, although the limit of III' being adiabatic is explored when $\beta$ is close to 1 and $\Delta_\mathrm{III'}\smap \Delta_\mathrm{I'}$.

\begin{figure*}[t]
   \centering
   \includegraphics[width=.85\textwidth]{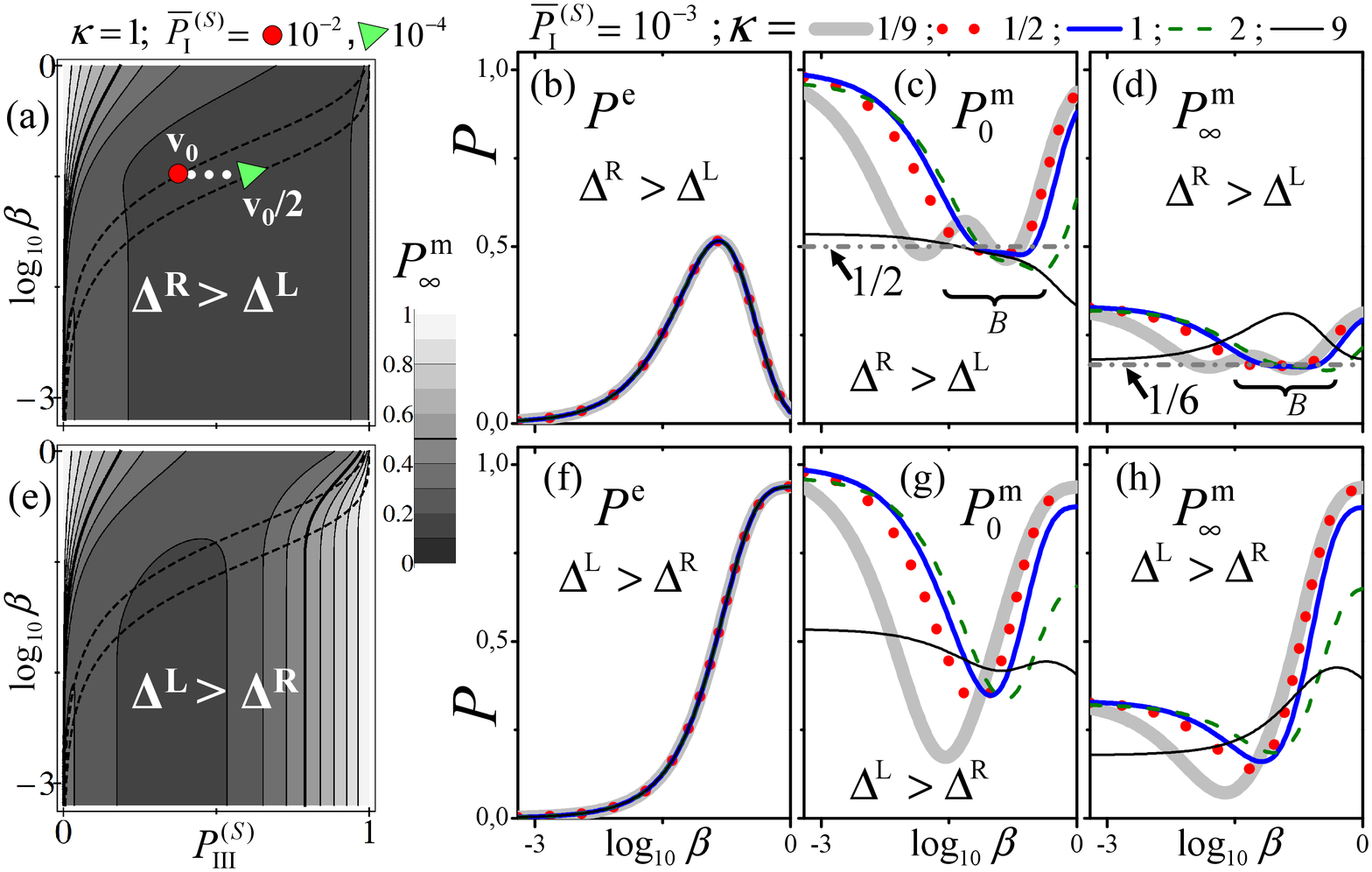}
  \vspace{-0.3cm}
\caption{Return probability for $\ketLR{\Psi_p}\smeq\ketLR{S^{(0,2)}_{++}}$ (see Fig\ref{FG:Pinf1}(b) and Fig.\ref{FG:detun} for the evolution and measurement regions, $R_\mathrm{e}$ and $R_\mathrm{m}$, respectively). Panels (a), (b), (c) and (d) are for $\Delta^\mathrm{L} \smlt \Delta^\mathrm{R}$ while panels (e), (f), (g) and (h) are for $\Delta^\mathrm{L} \smgt \Delta^\mathrm{R}$; see Fig.\ref{FG:cross}(c) and Eq.\eqref{EQ:MjMs}. Panels (a) and (e) show the saturation return probability $P_\infty^\mathrm{m}$ at the measurement region as a function \emph{of} $\beta$ and \emph{of} the LZ probability in the disorder induced crossing III', $P^{(\mathrm{s})}_\mathrm{III'}$, for $\kappa\smeq 1$. The two trajectories indicate the experimental conditions assuring that the LZ probability at crossing I' approaches the adiabatic limit, i.e., $\overline{P}^{(\mathrm{s})}_\mathrm{I'}$ is small. We choose $\overline{P}^{(\mathrm{s})}_\mathrm{I'}\smeq 10^{-3}$ and present the results as a function of $\beta$ for different values of $\kappa$ in the six panels (b) and (f), [probabilities at region $R^\mathrm{e}$, $P^\mathrm{e}$], (c) and (g) [probabilities at region $R^\mathrm{m}$ for $\tau_s\smeq 0$, $P_0^\mathrm{m}$], and (d) and (h) [probabilities at region $R^\mathrm{m}$ for $\tau_s\rightarrow \infty$, $P_\infty^\mathrm{m}$]. First, $P^\mathrm{e}$ is not zero and it depends on $\beta$, i.e., the system can remain in (1,1) after separation stage. Moreover, the ineffective separation stage combined with an ineffective joining stage make the return probabilities for $\tau_s\smeq 0$ \emph{smaller than $1$}. In panel (d) we find that the saturation return probability can be around 1/6 for a broad region in parameter space ($\beta,\kappa$), $B$; the associated $\tau_s\smeq0$ value in panel (c) is around 1/2. Results in (g) and (h) for $\Delta^\mathrm{L} \smgt \Delta^\mathrm{R}$ do not show the 1/6 saturation value with the same robustness as a function of $\beta$ and $\kappa$; such a difference could be tested by repeating the experiment preparing an (2,0) state (see text).}
   \label{FG:E6}
   \vspace{-0.4cm}
\end{figure*}

Once an excited state is prepared multiple Landau-Zener processes---including LZ loops as discussed in Sec.\ref{SC:MjMs}---affect the separation and joining stages. Furthermore, the Landau-Zener sequence of \emph{relevant} avoided crossings differs for $\DelR\smlt\DelL$ and $\DelL\smgt\DelR$. Obviously, these two cases can be investigated experimentally in the same device first by performing the RPE after the preparation of a desired excited state in the (0,2) configuration and then repeating the measurement preparing an equivalent excited state in the (2,0) charge configuration. In our calculations we always assume the preparation of an (0,2) state, therefore, the dot $\mathrm{R}$ referred to here must be identified with the dot in which the RPE is prepared with two electrons.

First we study the saturation return probability at the measurement region, $P^{\mathrm{m}}_\infty$, fixing $\kappa\smeq 1$.  For a given double dot $\beta$ is fixed and the separation and joining stages can be performed slower or faster changing the whole set of LZ-probabilities. With the assumptions discussed in Sec.\ref{SC:MjMs}, all the LZ-probabilities are linked once $\beta$ and $\kappa$ are given. In this case, we choose the LZ probability $P^{(\mathrm{s})}_\mathrm{III'}$ and $\beta$ as the two variables that define $P^{\mathrm{m}}_\infty$. We do so because here, as shown in Fig.\ref{FG:Pinf1}(b), the probability of state conversion at the avoided crossing III' plays a crucial role. The result is shown in Fig.\ref{FG:E6}(a) and (b) in the whole parameter space. In order to study situations experimentally relevant we assume that the detuning speed is tuned so that the avoided crossings with the biggest gaps are the adiabatic limit, i.e., $\overline{P}^{(\mathrm{s})}_{\mathrm{I'}}=P^ {(\mathrm{s})}_{\mathrm{leak}} \smap 0$. From Eq.\eqref{EQ:LZformula} and  Eq.\eqref{EQ:gapsBETAii} we find that the LZ probability of state conversion at III' is,
\be
P^{(\mathrm{s})}_{\mathrm{III'}}\left(\beta,P^{(\mathrm{s})}_{\mathrm{leak}}\right)= 1-\left(P^{(\mathrm{s})}_{\mathrm{leak}}\right)^{\beta}.
\label{EQ:PiiiBeta}
\ee
In Fig.\ref{FG:E6}(a) and (e) we show two curves with the values of $P^{(\mathrm{s})}_{\mathrm{III'}}$ resulting from choosing $P^{(\mathrm{s})}_{\mathrm{leak}}\smeq 10^{-2}$ or, as it results from halving the speed of the separation stage, $10^{-4}$.

We fix $P^{(\mathrm{s})}_{\mathrm{leak}}\smeq 10^{-3}$, (i.e., an intermediate situation between the latter two conditions) and we investigate the effectiveness of the separation stage by plotting $P^{\mathrm{e}}$ as a function of $\beta$; Figs.\ref{FG:E6}(c) and \ref{FG:E6}(f) correspond to $\DelR \smlt\DelL$ and $\DelR \smgt\DelL$, respectively. We can see that for $\beta\smlt 0.001$ the separation is effective, $P^{\mathrm{e}}\smap0$. The system is essentially \emph{clean} and only the DCAC I' is active. This is consistent with the result shown in Fig.\ref{FG:Mfac1}(b), for very small $\beta$ the LZ process in the DFAC can be neglected even though the LZ processes in the DCAC (visited with the same detuning speed) approach the adiabatic regime. This can also be seen from Eq.\eqref{EQ:PiiiBeta}, the small value of $P^{(\mathrm{s})}_{\mathrm{leak}}$---curves presented in Figs.\ref{FG:E6}(a) and \ref{FG:E6}(e) ---implies only for $\beta$ very small that $P^{(\mathrm{s})}_{\mathrm{III'}}$ can be neglected.

As $\beta$ grows up to 0.1 $P^{\mathrm{e}}$ also grows: the separation stage becomes ineffective. This results from a non-negligible $P^{(\mathrm{s})}_{\mathrm{III'}}$. However, for the range of $\beta$ between $0.1$ and $1$, $P^{\mathrm{e}}$ decreases for $\DelR \smlt\DelL$ and the separation stage improves, whereas for $\DelL \smlt\DelR$, $P^{\mathrm{e}}$ keeps growing with $\beta$. This is caused by the arrangements of the LZ processes, as sketched with gray arrows in the subpanels (i) and (iii) of Fig.\ref{FG:cross}(c). A finite $P^{(\mathrm{s})}_{\mathrm{III'}}$ implies the passage through the DCAC IV' in a detuning/energy configuration that differs depending on $\DelR$ and  $\DelL$. Due to the gap ratios in Eq.\eqref{EQ:gapsBETA} the LZ processes at IV' and IV are almost adiabatic (from $P^{(\mathrm{s})}_{\mathrm{leak}}\smeq 10^{-3}$ we get $P^{(\mathrm{s})}_{\mathrm{IV'}}\smap0.969$).  This means that for $\DelR \smlt\DelL$ the state  $\ketLR{S_{-,+}^{(1,1)}}$ is converted at IV' into $\ketLR{S_{-+}^{(0,2)}}$ and IV is irrelevant, whereas for $\DelL \smlt\DelR$ both IV' and IV are involved leading to the state $\ketLR{S_{+-}^{(1,1)}}$.

In the limiting case of $\beta\smeq 1$, the LZ processes in all the avoided crossings can be considered adiabatic. Since $P^{(\mathrm{s})}_{\mathrm{III}}\rightarrow 1$ the system follow $100\%$ the path indicated by the gray arrows in subpanels (i) and (iii) of Fig.\ref{FG:cross}(c). This means that for $\DelL \smlt\DelR$ the system leaves IV at state $\ketLR{S_{+-}^{(1,1)}}$ and then fully follows III (see Fig.\ref{FG:detun}) ending up in the $\ketLR{S_{--}^{(0,2)}}$ (0,2) state. This justifies the fact that for $\beta\smeq 1$ the separation stage is fully ineffective for $\DelL \smlt\DelR$ and $P^{\mathrm{e}}\rightarrow 1$. On the other hand, the separation stage is effective, $P^{\mathrm{e}}\rightarrow 0$, in the adiabatic limit for $\DelR \smlt\DelL$: the system leaves IV' at state $\ketLR{S_{+-}^{(0,2)}}$ and after the adiabatic state conversion at II' (see Fig.\ref{FG:detun}) it ends up in the $\ketLR{S_{++}^{(1,1)}}$ state.

It is clear, from Fig.\ref{FG:detun}, that only when the ground state is prepared the first avoided crossing involved in the separation stage is a doublet conserving one. Through the present example, we have shown a feature common to the case of preparing \emph{any} excited state of (0,2), namely, that if the region of detuning involving the avoided crossings is visited with \emph{a constant detuning speed} the separation stage is, in general, not effective. A return probability experiment performed in such a condition does not achieve the first conceptual goal of the RPE: separating the electrons. A detuning pulse engineered with different speeds at the different crossings can be the solution. This is the most important conclusion of this section.

For the sake of completeness, despite the ineffectiveness of the separation stage, we now show what return probabilities, $P^{\mathrm{m}}$, would be observed at the measurement region as a function of $\beta$ and fixing $P^{(\mathrm{s})}_{\mathrm{leak}}\smeq 10^{-3}$. The results are shown for different values of $\kappa$ in panels (c), (d), (g), and (h) of Fig.\ref{FG:E6}. Similarly to the case in the previous section, $\kappa\smeq 9$ implies a too fast joining stage that affects even the clean limit. For the smaller values of $\kappa$ the first observation is the trivial recover for $\beta\rightarrow 0$ of the $P^{\mathrm{m}}_0\smap 1$, $P^{\mathrm{m}}_\infty\smap 1/3$---consistent with the results with no valley mixing.\cite{ReynosoFlensberg2011} In a very disordered system with $\beta\rightarrow 1$ and $\DelR \smlt\DelL$, as discussed above, the separation stage is effective and again we recover the $P^{\mathrm{m}}_0\smap 1$, $P^{\mathrm{m}}_\infty\smap 1/3$ physics. On the other hand, for $\beta\rightarrow 1$ and $\DelL \smlt\DelR$ one measures $P^{\mathrm{m}}_0\smap 1$ and $P^{\mathrm{m}}_\infty\smap 1$, but the system does not have electrons separated in (1,1) states in the evolution stage.

Notably, for $\DelR \smlt\DelL$ we find an intermediate region in the $\beta$ axis, see zone $B$ in Fig.\ref{FG:E6}(d), for which $P^{\mathrm{m}}_\infty\smap 1/6$; the result persists even when the velocities of the separation and joining stages differ (for $\kappa\leq 2$). This value agrees very well with the saturation probability, 0.17, measured in Ref.~\onlinecite{ChurchillFlensberg2009}. However, while in the experiment for $\tau_s\rightarrow 0$ the return probability is 1, in our case study the ineffective separation and joining stages produce, in region $B$, $P^{\mathrm{m}}_0\smap 1/2$ as shown in Fig.\ref{FG:E6}(c).

Here, the robustness of the $P^{\mathrm{m}}_0\smap 1/2$, $P^{\mathrm{m}}_\infty \smap 1/6$ physics can be understood because the non-adiabatic LZ probabilities are linked by, $P^{(\mathrm{h})}_{\mathrm{II'}}\smeq 1- (1-P^{(\mathrm{h})}_{\mathrm{III'}})^{\frac{1}{2}}$ with $\mathrm{h}\smeq\{\mathrm{s},\mathrm{j}\}$. We find that the region $B$ approximately cover the values of $\beta$ that produce $0.3\smgt P^{(\mathrm{s})}_{\mathrm{III'}}\smgt 0.8$. We do not discuss further this situation because, as mentioned above, it is not a proper return probability experiment. However, a final interesting remark can be made regarding the experimental results of Ref.~\onlinecite{ChurchillFlensberg2009}. If the measured $P^{\mathrm{m}}_\infty \smap 1/6$  were due to the preparation of the highest excited state (0,2), as described here, then the preparation of the (2,0) analog state would not show such a saturation value with the same robustness as a function of $\kappa$ and $\beta$, this is shown in Fig.\ref{FG:E6}(h) with $\DelL \smlt\DelR$.

\subsection{RPE in the high magnetic field limit}
\begin{figure*}[!ht]
   \centering
   \includegraphics[width=.85\textwidth]{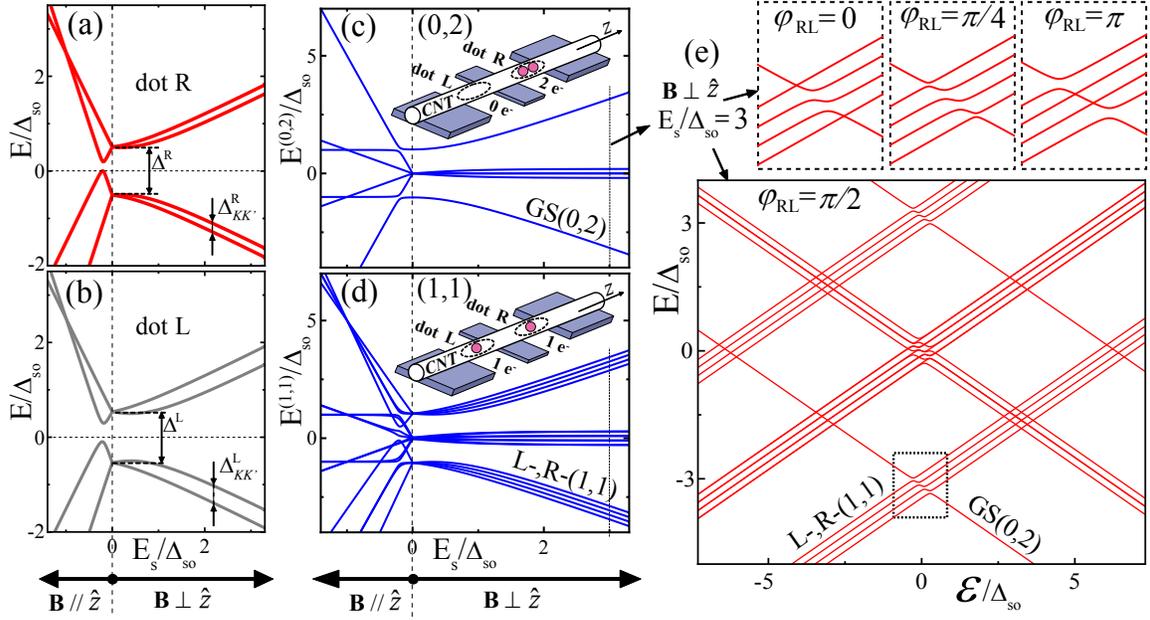}
\vspace{-0.3cm}
   \caption{Qualitative differences between parallel and perpendicular (to the tube axis) magnetic fields and their interplay with disorder. We take a common spin-orbit splitting $\Delta_\mathrm{so}$, and the disorder generates dot-dependent valley mixing energies, $\Delta^\mathrm{L}_\mathrm{\tiny KK'}\smneq\Delta^\mathrm{R}_\mathrm{\tiny KK'}$. Panels (a) and (b) show the single-particle eigenvalues for the isolated right and the left dot, respectively, as a function of the Zeeman energy $E_\mathrm{s}$; we distinguish the parallel from the perpendicular magnetic field case. At the high-field limit, $\Delta_\mathrm{so}\ll|E_s|$, the valley mixing effect is irrelevant for the parallel case (due to the diamagnetic term) while on the other hand it is important for the perpendicular case. It follows that in the high-field limit parallel magnetic fields leads to the same physics as in a clean nanotube.\cite{ReynosoFlensberg2011} In panels (c) and (d) we plot the energies of the (0,2) and (1,1) eigenstates (in the high detuning limit or for $t\smeq 0$). Differences in the two valley mixing energies generate splitting of the four lowest energy (1,1) states for perpendicular magnetic field case. In panel (e) we plot the mixing between (0,2) and (1,1) states as a function of the detuning for a case in which the Zeeman energy due to the perpendicular magnetic field is dominant. We focus on the mixing involving the (0,2) ground state, GS(0,2). The (0,2) ground state is spin polarized and therefore can only mix with the four lowest energy states of (1,1), $\mathrm{L-,R-}$, that are also spin polarized. As follows from Eq.\eqref{EQ:tunB} the tunneling is determined by the valley properties and therefore the phase difference between the valley mixings, $\varphi_{RL}$, tunes the mixing. In the insets above we show that, unless $\varphi_{LR}\smeq n\pi$, one should expect mixing of GS(0,2) with \emph{the four} $\mathrm{L-,R-}$ states of the (1,1) configuration.}
   \label{FG:finiteB}
\vspace{-0.3cm}
\end{figure*}

In spin-only double quantum dots one of the most common procedures is applying an external magnetic field much larger than the hyperfine interaction energy. Focusing on the (1,1) states, the Zeeman interaction shifts in energy the spin polarized triplets leaving in a double degenerated subspace the spin-singlet state $\ketLR{S}\smeq\ketLR{\uparrow\downarrow}\smmi\ketLR{\downarrow\uparrow}$ and the triplet-0 state $\ketLR{T_0}\smeq\ketLR{\uparrow\downarrow}\smpl\ketLR{\downarrow\uparrow}$. This two level system defines a qubit that can be manipulated. The return probability experiment becomes simpler than for the case for zero magnetic field. The (0,2) singlet is prepared and after the separation stage the inhomogeneous part of the hyperfine interaction mixes the (1,1) singlet with the triplet-0. The result of averaging over many realizations leads to $P_\infty\smeq1/2$ in the high detuning limit (negligible tunneling exchange).\cite{CoishDQD,Petta2005,TaylorDephasingTheory2007}

The search for two-level subspaces that can be manipulated naturally leads to the investigation of the effect of external magnetic field. Here we focus on how the return probability behaves in such cases for double dots in nanotubes with valley mixing. The finite field case was studied for the clean nanotube system in Ref.~\onlinecite{ReynosoFlensberg2011}. Here, our goal is exploring the high magnetic field case, i.e., when the magnetic energies are much bigger than the hyperfine energy, the spin-orbit splitting and the valley mixing energies. The form of the single-particle solutions at those limits determines the nature of the (0,2) and (1,1) solutions and the behavior of the experiment.

We solve the dot Hamiltonian of Eq.\eqref{EQ:Hsingledot}, in Fig.\ref{FG:finiteB}(a) and (b) we plot the single-particle energies as a function of the Zeeman energy $E_\mathrm{s}$ for the isolated left and right dot assuming they have different valley mixing energies. We present in the same panels the cases with ${\bf B}\parallel\zver$ and ${\bf B}\perp\zver$ where $\zver$ is taken along the tube axis. The two particle energies for the (0,2) and the (1,1) charge configurations (at $t\smeq0$ or in high detuning limit) are plotted in panels (c) and (d) of Fig.\ref{FG:finiteB}; they follow from Eq.\eqref{EQ:02sols0En} and Eq.\eqref{EQ:11sols0}.

We first focus on the ${\bf B}\parallel\zver$ case. Neglecting the HFI, the spin projection along the $z$-axis, $\sigma$, is a good quantum number. The general single-dot single-particle solutions for this case is given in Eq.\eqref{EQ:eigen}. The solutions in valley space are the spinors $\ketLR{\hat{\varsigma}^{\xi}_{\sigma},\pm}$, they point parallel (+) and antiparallel (-) to the valley vector $\hat{\varsigma}^{\xi}_{\sigma}$. In the high-field limit the valley vector component \emph{out-of-the-plane} dominates due to the diamagnetic component that couples with the valley degree of freedom, $|\delta_{\sigma}^{\xi}|\smeq |\sigma\Delta_{so}^{\xi} -  E_{\rm orb}|\smgg  \Delta_{\tiny KK'}^{\xi}$, and the valley spinors $\ketLR{\hat{\varsigma}^{\xi}_{\sigma},\pm}$ become $\ketLR{\tau}$, with $\tau\smeq K,K'$ (the $\pm1$ eigenstates of $\tau_3$). This means that the solutions for the dots are simply $\ketLR{\xi\tau \sigma}$ with eigenenergies,
\begin{equation}
E_{\xi\tau\sigma}= \left(E_{\mathrm{s}}\sigma + E_{\mathrm{orb}}\tau - \Delta_{\mathrm{so}} \tau\sigma  \right)/2, \label{EQ:ENER1p1d}
\end{equation}
where $\tau$ is $+1$ or $-1$ for $K$ or $K'$, respectively.
This result is identical to a clean nanotube quantum dot with spin-orbit coupling in a parallel field. This is confirmed by the (0,2) and (1,1) spectrum at the high ${\bf B}\parallel\zver$ limit in Fig.\ref{FG:finiteB}(c) and (d). The available return probability situations correspond to those reported in Ref.~\onlinecite{ReynosoFlensberg2011}, leading to $P_\infty\smeq 1/2$ or $P_\infty\smeq 1$ depending on the prepared state. No disorder-induced avoided crossings survive in such a limit because the solutions in both dots have the same spin and valley characteristics. Any given (0,2) state is mixed, due to interdot tunneling, with a single (1,1) state and the gap energy is $2\sqrt{2}|t|$ in all six avoided crossings.

We now focus on the perpendicular magnetic field case, we take ${\bf B}\smeq B_x \xver$. We choose the single particle basis $\ketLR{\xi
\tau s_x}$ with $s_x\!=\uparrow_x,\downarrow_x$ (or $s_x\smeq \pm$) the spin projection along the direction of the magnetic field.  The spin-orbit coupling term in the Hamiltonian of Eq.\eqref{EQ:Hsingledot} is $-\frac{1}{2} \Delta_{so}^{\xi} \tau_3 \sigma_z$ and mixes $\ketLR{\xi
K s_x}$ with $\ketLR{\xi
K \bar{s_x}}$. Such a mixing can be neglected in the high magnetic field limit because states with opposite spin projections $s_x$ are split in energy by the Zeeman term, $E_{s} \tau_0 \sigma_x$. However, the valley mixing term in the Hamiltonian mixes
states of opposite valley but identical spin projection $s_x$; the effective Hamiltonian at high $B_x$ for the electrons $s_x$ in dot $\xi$ is the valley operator,
\be
H^{\perp}_{\xi,s_x}=\frac{1}{2}\left(\begin{array}{cc}  E_{s} s_x & \Delta_{\tiny KK'}^{\xi} \exp\left(\ci \varphi_{\tiny KK'}^{\xi} \right) \\
\Delta_{\tiny KK'}^{\xi} \exp\left(-\ci \varphi_{\tiny KK'}^{\xi} \right) & E_{s} s_x
\end{array}\right).
\ee
The solutions are the in-plane valley spinors,
\be
\ketLR{u_\mathrm{\xi}}\equiv\frac{1}{\sqrt{2}}\left( \begin{array}{c}
{\rm e}^{\frac{\ci}{2} \varphi_{\tiny KK'}^{\xi}} \\
\pm{\rm e}^{-\frac{\ci}{2} \varphi_{\tiny KK'}^{\xi}}
\end{array}
\right),
\ee
with $u_\mathrm{\xi}\smeq\pm$. Thus, the full single-particle solutions are $\ketLR{\xi u_\mathrm{\xi} s_x} = \ketLR{\xi} \otimes\ketLR{u_\mathrm{\xi}}\otimes \ketLR{s_x}$, with energies,
\begin{equation}
E_{\xi u_\mathrm{\xi} s_x}= \left(E_{\mathrm{s}} s_x + u_\mathrm{\xi} \Delta_{\tiny KK'}^{\xi}  \right)/2. \label{EQ:ENER1p1d2}
\end{equation}

The four (1,1) lowest energy states Fig.\ref{FG:finiteB}(d) are the (1,1) four Slater determinants with $s_x\smeq \downarrow_x$, i.e., $\ketLR{{}^{\mathrm{L} u_\mathrm{L} \downarrow_x}_{\mathrm{R} u_\mathrm{R} \downarrow_x}}$ with $u_\mathrm{R}\smeq\pm$ and $u_\mathrm{L}\smeq\pm$. In general, $\Delta_{\tiny KK'}^{\mathrm{R}}\smneq\Delta_{\tiny KK'}^{\mathrm{L}}$ and therefore the valley mixing breaks the degeneracy between these four states because (disregarding a global energy shift) the energies are,
\be
E^{(1,1),\downarrow_x}_{u_\mathrm{L},u_\mathrm{R}}= -E_{\mathrm{s}}  + \frac{u_\mathrm{R}}{2} \Delta_{\tiny KK'}^{\mathrm{R}} + \frac{u_\mathrm{L}}{2} \Delta_{\tiny KK'}^{\mathrm{L}}.
\ee
This is a clear distinction of this system with spin-only double dots, there is not any double degenerated subspace (analogous to the one spanned by $\ketLR{T_0}$ and $\ketLR{S}$) in the (1,1) configuration.

The last issue we investigate for perpendicular field is how the (0,2) ground state, $\ketLR{{}^{\mathrm{R} + \downarrow_x}_{\mathrm{R} - \downarrow_x}}$, mixes with the (1,1) states. This is important for the return probability experiment if the ground state is to be prepared. First due to spin conservation of the single particle tunneling, $H_T\smeq -t \xi_1 \tau_0 \sigma_0$, the two spin-down electrons can only mix with the four lowest energy (1,1) states having the same spin configuration. The result is confirmed in Fig.\ref{FG:finiteB}(e) where we plot the two particle spectrum as a function of detuning choosing a high value of $B_x$. We notice that the phase difference between the valley mixing interactions in the two dots, $\varphi_{RL} \smeq \varphi_{\tiny KK'}^{R}\smmi \varphi_{\tiny KK'}^{L}$, plays a key role in weighting the tunneling between the (0,2) ground state and the different (1,1) states. This is shown in the insets in Fig.\ref{FG:finiteB}(e).

Such a dependence follows from computing the matrix elements of the tunneling Hamiltonian between solutions with a given spin in different dots, we get:
\be
\braLR{\mathrm{L} u_{\mathrm{L}} s_x} H_T \ketLR{\mathrm{R} u_{\mathrm{R}} s_x}= \begin{cases}
-t\cos\frac{\varphi_{RL}}{2} & \text{for $u_{\mathrm{R}} \smeq u_{\mathrm{L}}\smeq \pm$}  \\
-t\ci \sin\frac{\varphi_{RL}}{2} & \text{for $u_{\mathrm{R}} \smeq -u_{\mathrm{L}} \smeq \pm$}
\end{cases}.
\label{EQ:tunB}
\ee
Then the application of the tunneling Hamiltonian to the (0,2) ground state gives ,
\bea
H_{T}^{\rm 2p} \ketLR{{}^{\mathrm{R} + \downarrow_x}_{\mathrm{R} - \downarrow_x}} &=& -t \cos\frac{\varphi_{RL}}{2} \left( \ketLR{{}^{\mathrm{L} + \downarrow_x}_{\mathrm{R} - \downarrow_x}} +\ketLR{{}^{\mathrm{R} + \downarrow_x}_{\mathrm{L} - \downarrow_x}}\right)\nonumber \\&&-t \ci \sin\frac{\varphi_{RL}}{2} \left( \ketLR{{}^{\mathrm{L} - \downarrow_x}_{\mathrm{R} - \downarrow_x}} +\ketLR{{}^{\mathrm{R} + \downarrow_x}_{\mathrm{L} + \downarrow_x}}\right).
\eea
Therefore, for $\varphi_{RL}\smeq 0$, $\sin\frac{\varphi_{RL}}{2}\smeq 0$ and the $(0,2)$  ground state mixes only with the two intermediate energy states, while on the other hand for $\varphi_{RL}\smeq \pi$, $\cos\frac{\varphi_{RL}}{2}\smeq 0$ and the mixing is only with the other two states. For other values of $\varphi_{RL}$ there are tunneling amplitudes with the four states.

All these effects have to be considered if a return probability experiment is to be performed preparing the (0,2) ground state. We note that if $\frac{1}{2}\left|\Delta_{\tiny KK'}^{\mathrm{R}}-\Delta_{\tiny KK'}^{\mathrm{L}}\right| \gg \sigma_\mathrm{H}$, the hyperfine interaction would not induce mixing between the (1,1) states. However, the experiment, if performed with different detuning speeds and different waiting times $\tau_s$, can provide information about the values of $\Delta_{\tiny KK'}^{\mathrm{\xi}}$ and $\varphi_{RL}$. Some oscillations in the signal would not average out when repeating the single-shot cycle if the values of $\Delta_{\tiny KK'}^{\mathrm{\xi}}$ are fixed in time. This would be the case if its origin is indeed static disorder and defects.
\section{Conclusions}
The return probability experiment is a standard tool for characterizing dephasing in double quantum dots systems. The valley degree of freedom in nanotubes introduces more states in the system. The spin-orbit coupling, the Zeeman interaction, and the diamagnetic effect of a parallel magnetic field, are elements to be considered when performing the experiment. In ultra clean nanotube quantum dots, in which one can neglect valley mixing effects, the return probability experiment presents many different scenarios.\cite{ReynosoFlensberg2011} In such a case, the origin of the different cases lies in the effective hyperfine induced dynamics within subspaces of (1,1) states. Those subspaces are defined by the chosen parameters. On the other hand, the separation and joining stages of the experiment are trivial.

In this paper we focus on nanotubes for which the valley mixing energy is much larger than the hyperfine energy, $\sigma_\mathrm{H}$, and therefore it cannot be neglected. The presence of valley mixing could be due to defects and impurities in the nanotube. Since the disorder profile is position dependent the resulting valley mixing terms are dot-dependent. First, we focused on the zero-magnetic field case and solved the single-dot problem. At the dot $\xi\smeq\{ \mathrm{L,R}\}$ the spin-orbit coupling and the valley mixing split the four states, $\ketLR{\xi \mathrm{d}_\xi \sigma}$, an energy $ \mathrm{d}_\xi \Delta^{\xi}/2$ into two Kramers doublets, the low energy one, $\mathrm{d}_\xi\smeq -$ and the high energy one $\mathrm{d}_\xi\smeq +$. Because $\sigma_\mathrm{H}\ll \Delta^{\xi}$ the hyperfine interaction is unable to mix states from different doublets. We have shown that the effective hyperfine interaction in the space of the two states in a Kramers doublet remains statistically isotropic as for the case of clean nanotubes. The dynamics due to hyperfine interaction in the (1,1) states is, as in spin-only double dots, dictated by precession along the local hyperfine effective fields that are different in the two dots and in the two Kramers doublets.

In the high detuning limit in absence of hyperfine interaction the sixteen (1,1) eigenenergies split into the four fourfold degenerated subsets, $(\mathrm{L} \mathrm{d_L},\mathrm{R} \mathrm{d_R})$. The action of the hyperfine interaction cannot change the Kramers doublet index $\mathrm{d}_\xi$ and thus the final state lies in the same subset as the initial state. We get the standard behavior for the hyperfine dynamics within each subset: if starting in a singlet-like state, the probability to be found at $\tau_s\gg\hbar/\sigma_\mathrm{H}$ in the same state saturates (averaging over many single-shot experiments) to a value $1/3$. It is worth noting that the isotropic hyperfine field we have obtained is the most effective coherent form for dephasing the system. This isotropy, in average, leads to the lowest possible saturation value; the presence of anisotropy in the hyperfine field---which can arise in clean systems for perpendicular magnetic fields---increases the saturation value up to $3/8$.\cite{ReynosoFlensberg2011} The situation we are studying is therefore the most favorable to achieve a secondary goal, obtaining the lowest possible saturation values allowed in the system given that the only experimental result available is lower than $1/3$.\cite{ChurchillFlensberg2009}

While the valley mixing does not introduce qualitative changes in the hyperfine interaction it does so for the single-particle interdot tunneling. The solutions in each dot point in noncolinear directions in valley space. This is shown to produce nonzero tunneling matrix elements between solutions in different dots with the same spin $\sigma$ \emph{irrespective} of the Kramers doublet index. In the two particle spectrum reflecting the mixing between the (1,1) and (0,2) states as a function of detuning, we show that several new avoided crossings appear as a consequence of the doublet-flipping interdot tunneling amplitudes. These avoided crossings can play a crucial role in the separation and joining stages modifying the output of the experiment.

For gaining qualitative knowledge of the implications of the novel crossings we adopt a low tunneling picture that allows us to use the Landau-Zener formula for each avoiding crossing. The parameter $\beta$, i.e., the absolute square ratio between the doublet-flipping and the doublet-conserving tunneling amplitudes, defines the relations among all the gaps in the avoided crossings. We show that when any excited (0,2) state is prepared the separation stage first involves an avoided crossing induced by disorder. Any separation stage that passes through the mixing region \emph{with the same detuning speed} (a non-fine-tuned detuning pulse) can lead to imperfect separation leaving the electrons in the same dot. This is not consistent with a proper return probability experiment since it leads to $P(\tau_s\smeq 0)\smlt 1$; for example we show that when preparing the highest excited state one gets $P(\tau_s\smeq 0)\smap 1/2$ and $P_\infty\smap 1/6$ in a broad region in parameter space. Despite the agreement of the saturation value with the experimental one,\cite{ChurchillFlensberg2009} 0.17, there exists a strong discrepancy at short evolving times because the measurement goes to 1. As mentioned above more experimental work would be desirable, including the dependence with magnetic fields.\cite{ReynosoFlensberg2011}

We also showed that preparing the ground state leads to a separation stage that first involves the doublet conserving avoided crossing I. This crossing can be visited adiabatically and no further avoided crossings are involved in the separation stage. We find that the return probability experiment behaves very similar to the spin-only case, except that the $(1,1)$ triplet-like states (with $2/3$ probabilities of being occupied in average at $\tau_s\rightarrow \infty$) have a nonzero probability, $P^{(\mathrm{j})}_{\mathrm{II}}$, of \emph{returning} to (0,2) due to the disorder induced LZ-process at the avoided crossing II. For the saturation return probability this implies that it cannot be smaller than $1/3$, namely, $P_\infty\smap 1/3 \smpl 2/3 P^{(\mathrm{j})}_{\mathrm{II}}$.

Finally, we have studied the high magnetic field limit. When the magnetic field is parallel to the tube axis the diamagnetic effect competes directly with the valley mixing term because both operate in valley space for electrons with the same spin projection. This makes the valley mixing irrelevant for high magnetic fields, the single particle solutions become valley polarized and the return probability experiment produces the same results predicted for clean nanotubes in parallel magnetic fields.\cite{ReynosoFlensberg2011} On the other hand, if the magnetic field is perpendicular to the tube axis there is no diamagnetic interaction and the valley mixing splits states with the same spin that otherwise would be degenerated at high magnetic fields. We show that the phase difference between the valley mixing terms in the two dots controls how the interdot tunneling behaves and this changes the avoided crossing pattern between the (0,2) ground state and the (1,1) four lowest energy states (all being fully spin-polarized). Though the hyperfine interaction would play no role, variations of the return probability experiment could give insight of the valley mixing terms by studying the oscillations in the signal. The combination of the multiple mixing of the (0,2) ground state with the fact that no double degenerate (1,1) subspace remains due to disorder---analogous to the singlet and triplet-0 subspace in spin-only double dots---can be important in the interpretation of Pauli blockade measurements in nanotubes in a perpendicular magnetic field.\cite{Buitelaar2011}

\acknowledgments We acknowledge useful discussions with G. Burkard, H. Churchill, K. Grove-Rasmussen, F. Kuemmeth, C. Marcus, D.J. Reilly, M. Wardrop and S. Weiss. AAR acknowledges support from the Niels Bohr International Academy, the Australian Research Council Centre of Excellence scheme CE110001013 and from ARO/IARPA project W911NF-10-1-0330. KF acknowledges support from the Danish Council for Independent Research. KF is grateful for the hospitality of The Quantum Science Group at the University of Sydney.

%\bibliographystyle{apsrev4-1}
%\bibliography{CNTbase112011}
%merlin.mbs apsrev4-1.bst 2010-07-25 4.21a (PWD, AO, DPC) hacked
%Control: key (0)
%Control: author (72) initials jnrlst
%Control: editor formatted (1) identically to author
%Control: production of article title (-1) disabled
%Control: page (0) single
%Control: year (1) truncated
%Control: production of eprint (0) enabled
%

\end{document}